\documentclass[journal]{journal}
\pdfoutput=1
\usepackage{multicol,lipsum}
\usepackage{booktabs}
\usepackage[normalem]{ulem}
\usepackage{longtable}
\useunder{\uline}{\ul}{}

\ifCLASSINFOpdf
	\usepackage[pdftex]{graphicx}
 
\else
\fi
\usepackage[cmex10]{amsmath}

\usepackage{mathtools}

\usepackage{float}

\usepackage[export]{adjustbox}

\usepackage{fixltx2e}
\usepackage{ulem}
\usepackage{subfigure}
\usepackage{xcolor}
\usepackage{setspace}

\usepackage[colorlinks,
            linkcolor=blue,
            anchorcolor=blue,
            citecolor=blue]{hyperref}

\usepackage{nameref}

\begin{document}
%
\title{Newtonian Mechanics Based Transient Stability PART V: Inner-group Machine}
%
%
%

\author{{Songyan Wang,
        Jilai Yu,  
				Aoife Foley,
        Jingrui Zhang
        }
        
}
%
%

\markboth{Journal of \LaTeX\ Class Files,~Vol.~6, No.~1, January~2007}%
{Shell \MakeLowercase{\textit{et al.}}: Bare Demo of IEEEtran.cls for Journals}
%



\maketitle
\thispagestyle{empty}
\begin{abstract}
This paper analyzes the mechanisms of the inner-group machine. It is first clarified that the inner-group machine is created from the difference between the equivalent system and the original system. The inner-group machine stability is analyzed based on the machine paradigms. In particular, strict correlation between the inner-group machine trajectory and the inner-group machine transient energy conversion is established through the I-CR system modeling. Then, the transient characteristics of the inner-group machine are analyzed. It is clarified that the inner-group motions might be inseparable or separable, and the inner-group machine DLP will occur later than the EDLP and IDLP. Simulation results show that the severity of the original system cannot be simply evaluated through its equivalent system once any inner-group motion becomes fierce.
\end{abstract}

\begin{IEEEkeywords}
Transient stability, transient energy, equal area criterion, individual machine, inner-group motion
\end{IEEEkeywords}
%

\IEEEpeerreviewmaketitle
\begin{small}

\begin{tabular}{lllll}
    &            &               &                  &                                \\
  \multicolumn{5}{c}{\textbf{Nomenclature}}                                                \\

  DLP    &  \multicolumn{1}{c}{} & \multicolumn{3}{l}{Dynamic liberation point}            \\
  DSP    &                       & \multicolumn{3}{l}{Dynamic stationary point}            \\
  EAC    &                       & \multicolumn{3}{l}{Equal area criterion}                \\
  TSA    &                       & \multicolumn{3}{l}{Transient stability assessment}      \\
  MOD    &                       & \multicolumn{3}{l}{Mode of disturbance}      \\
  COI    &                       & \multicolumn{3}{l}{Center of inertia}      \\
  GTE    &                       & \multicolumn{3}{l}{Global total transient energy}           \\
  IVCS    &                       & \multicolumn{3}{l}{Individual-machine virtual COI-SYS machine}           \\
  LOSP    &                       & \multicolumn{3}{l}{Loss of synchronism point}           \\ 
  IMTR    &                       & \multicolumn{3}{l}{Individual-machine trajectory}           \\ 

\end{tabular}
\end{small}

\section{Introduction} \label{section_I}

%
%
%
%
\subsection{LITERATURE REVIEW} \label{section_IA}
\raggedbottom
\IEEEPARstart{T}{he} inner-group machine motions can be seen as the differences between the original system and the equivalent system \cite{1}-\cite{3}.
It reflects the relative motion between the individual machines and the equivalent machine inside each group. Although the stabilities of the original system and the equivalent system are identical \cite{3}, the severities of the two systems might be different once any inner-group machine motion becomes fierce.
In particular, the severities of the two systems will be close if all the inner-group motions are slight. The severities of the two systems will be different if any inner-group motion is fierce. This also indicates that the original system cannot be completely replaced with the equivalent system under certain distinctive circumstances.
\par The inner-group machine motion is ``created'' from the difference between the original system and the equivalent system. In other words, it does not exist in the original system and neither can be found in the equivalent system. The discussions about the inner-group motions were only very a few in the history of the equivalent machine studies. Xue stated that the stability margin of the system might show great errors if the inner-group-machine motion cannot be neglected, especially when the group separation pattern is not clear \cite{4}.
In fact, the precise modeling of the inner-group motion was rarely analyzed even in the modern equivalent machine studies \cite{5}. Against this background, the explorations of the inner-group machine stability will become of value, because it may essentially clarify the complicated relationship between the individual-machine and the equivalent machine that both follow the machine paradigms and show effectiveness in TSA.

\subsection{SCOPE AND CONTRIBUTION OF THE PAPER} \label{section_IB}
Following the tutorial explanation of the inner-group motions as given in the previous paper \cite{3}, in this paper the authors focus on the modeling and stability characterizations of the inner-group machine motions. The mechanism of the inner-group machine is analyzed.
It shows that the inner-group machine can be defined as the ``difference'' between the original multi-machine system and the equivalent system. Based on the machine paradigms, the trajectory variance of the inner-group machine in the COI-CR reference is modeled through the I-CR system with strict NEC characteristic. Then, the transient characteristics of the inner-group machine are analyzed. The inner-group motions might be inseparable or separable, and the inner-group instability will occur when the inner-group machine separates from Machine-CR.
In the end of the paper, the ``companion'' relationship between the individual-machine and the equivalent machine in TSA are analyzed. It is found that the equivalent machine shows flexibility when the inner-group machine motions are slight, while the individual-machine can be used in certain distinctive situations in which ``more efficient stability characterization'' or ``more precise severity evaluation'' becomes a particular emphasis.
\par The contributions of this paper are summarized as follows
\\ (i) It is clarified that the inner-group machine motion is created from the difference between the original multi-machine system and the equivalent system. This reveals the mechanism of the inner-group motions.
\\ (ii) The inner-group machine stability is analyzed based on the machine paradigms. This provides a precise modeling and stability characterization for the inner-group machine.
\\ (iii) The effect of the inner-group machine instability to the original system stability is analyzed. This clarifies that the original system cannot be simply replaced with the equivalent system in TSA especially when any inner-group machine instability occurs.
\par The reminder of the paper is organized as follows. In Section \ref{section_II}, the inner-group machine stability is analyzed based on the transient stability paradigms. In Section \ref{section_III}, the transient characteristics of the inner-group machine are analyzed. In Section \ref{section_IV}, simulation cases are given to demonstrate the inner-group machine stability.
In Section \ref{section_V}, the companion relationship between the individual-machine and the equivalent machine in TSA is analyzed. In Section \ref{section_VI}, comparison between MOD and group separation pattern is analyzed. In Section \ref{section_VII}, the machines in the previous papers are systemically revisited. Conclusions are given in Section \ref{section_VIII}.
\par In this paper, all the analysis about the inner-group motion is given under the COI-SYS reference. The CR-NCR system is replaced with its mirror systems, i.e., the CR-SYS system and the NCR-SYS system.

\section{PARADIGMS BASED INNER-GROUP MACHINE ANALYSIS} \label{section_II}
\subsection{CREATIONS OF THE INNER-GROUP MOTIONS}  \label{section_IIA}

In the previous paper \cite{3}, the inner-group motions are analyzed in tutorial under dominant group separation pattern. In this section, it will be extended to the environment with all the possible group separation patterns.
\par
\begin{figure}[H]
  \centering
  \includegraphics[width=0.45\textwidth,center]{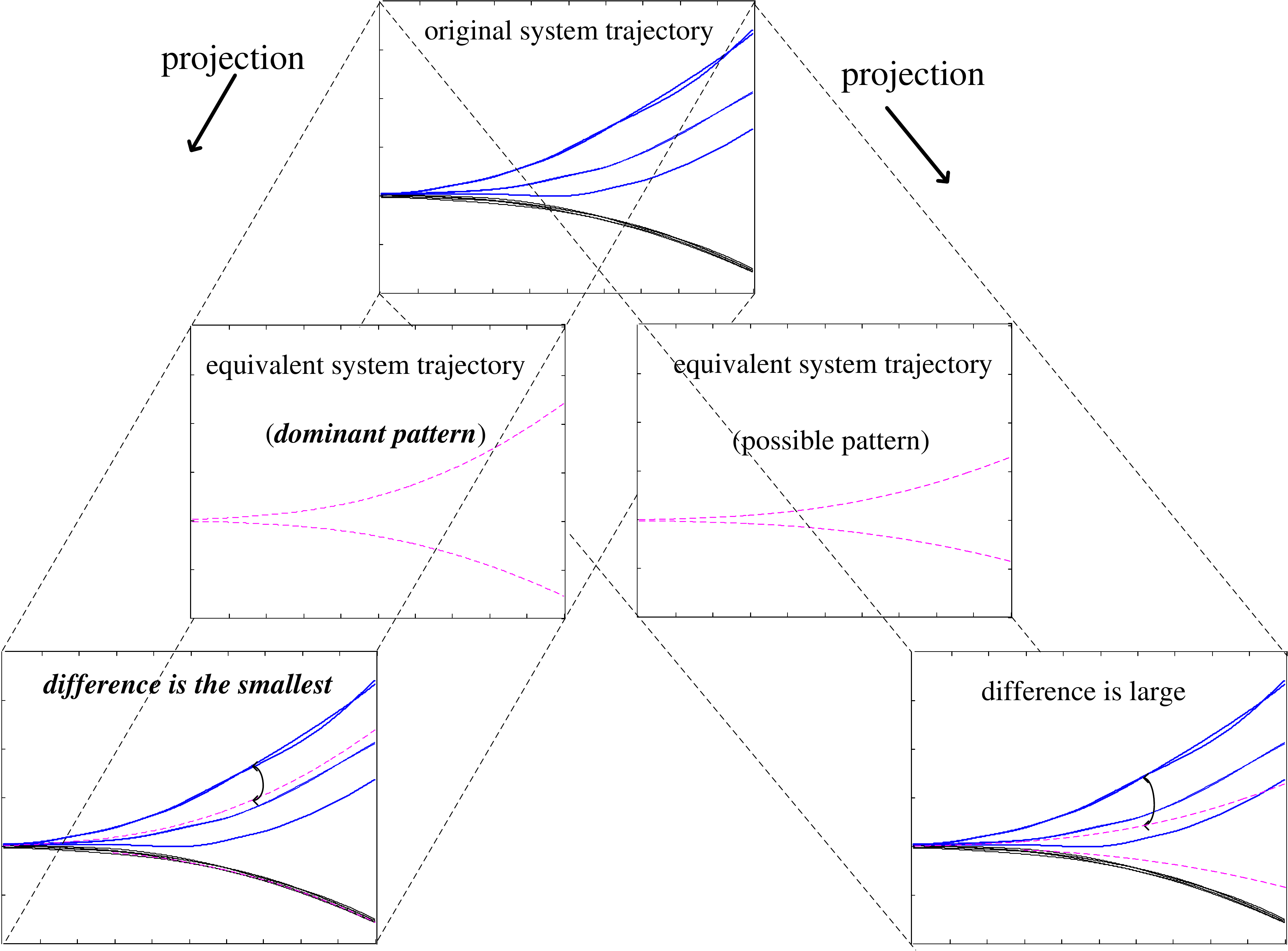}
  \caption{Inner-group motions under each possible group separation pattern.} 
  \label{fig1}  
\end{figure}

 Assume projections can be established through original system trajectory and the equivalent system trajectory in each group separation pattern. The equivalent system trajectory under each group separation patterns is set as a ``slice'' for the projection. Demonstration about this projection is shown in Fig. \ref{fig1}.
\vspace*{-0.5em}
From Fig. \ref{fig1}, the inner-group machine motions are created from the difference between the original system and the equivalent system in each group separation pattern. Further, since the equivalent system under the dominant pattern is the closest to the original system, the inner-group machine motions under the dominant pattern will be the slightest among the cases in all the possible patterns.
\par The inner-group motions do not exist in the equivalent system, and neither can be found in the equivalent system. It reflects the separation of an individual-machine with respect to the equivalent machine inside the group. In other words, the inner-group motion reflects the relative motion of the ``inner-group machine'' by using the equivalent machine as the motion reference. Against this background, the inner-group machine stability can be naturally analyzed through machine paradigms.

\subsection{INNER-GROUP MACHINE MONITORING}  \label{section_IIB}
In this case we take the inner-group motions inside Group-CR as an example. The analysis can be simply extended to the case in Group-NCR. Note that all the analysis in the following paper is based on the dominant group separation pattern.
\par The equation of motion of each individual machine in the COI-SYS reference is denoted as 
\begin{equation}
  \label{equ1}
  \left\{\begin{array}{l}
    \frac{d \delta_{i\mbox{-}\mathrm{SYS}}}{d t}=\omega_{i\mbox{-}\mathrm{SYS}} \\
    \\
    M_{i} \frac{d \omega_{i\mbox{-}\mathrm{SYS}}}{d t}=f_{i\mbox{-}\mathrm{SYS}}
    \end{array}\right.
\end{equation}
\par In Eq. (\ref{equ1}), all the parameters are given in Ref. \cite{1}.
\par Then, the equivalent Machine-CR is used as the RM. The equation of motion of Machine-CR in the COI-SYS reference is depicted as
\begin{equation}
  \label{equ2}
  \left\{\begin{array}{l}
    \frac{d \delta_{\mathrm{CR}\mbox{-}\mathrm{SYS}}}{d t}=\omega_{\mathrm{CR}\mbox{-}\mathrm{SYS}} \\
    \\
    M_{\mathrm{CR}} \frac{d \omega_{\mathrm{CR}\mbox{-}\mathrm{SYS}}}{d t}=f_{\mathrm{CR}\mbox{-}\mathrm{SYS}}
    \end{array}\right.
\end{equation}
\par In Eq. (\ref{equ2}), all the parameters are given in Ref. \cite{1}.
\par The inner-group-machine trajectory (IGMTR) is denoted as
\begin{equation}
  \label{equ3}
  \delta_{i\mbox{-}\mathrm{CR}}=\delta_{i}-\delta_{\mathrm{CR}}=\delta_{i\mbox{-}\mathrm{SYS}}-\delta_{\mathrm{CR}\mbox{-}\mathrm{SYS}}
\end{equation}
\par From Eq. (\ref{equ3}), the IGMTR depicts the separation of the machines inside the group with respect to Machine-CR.
\par
\begin{figure}[H]
  \centering
  \includegraphics[width=0.38\textwidth,center]{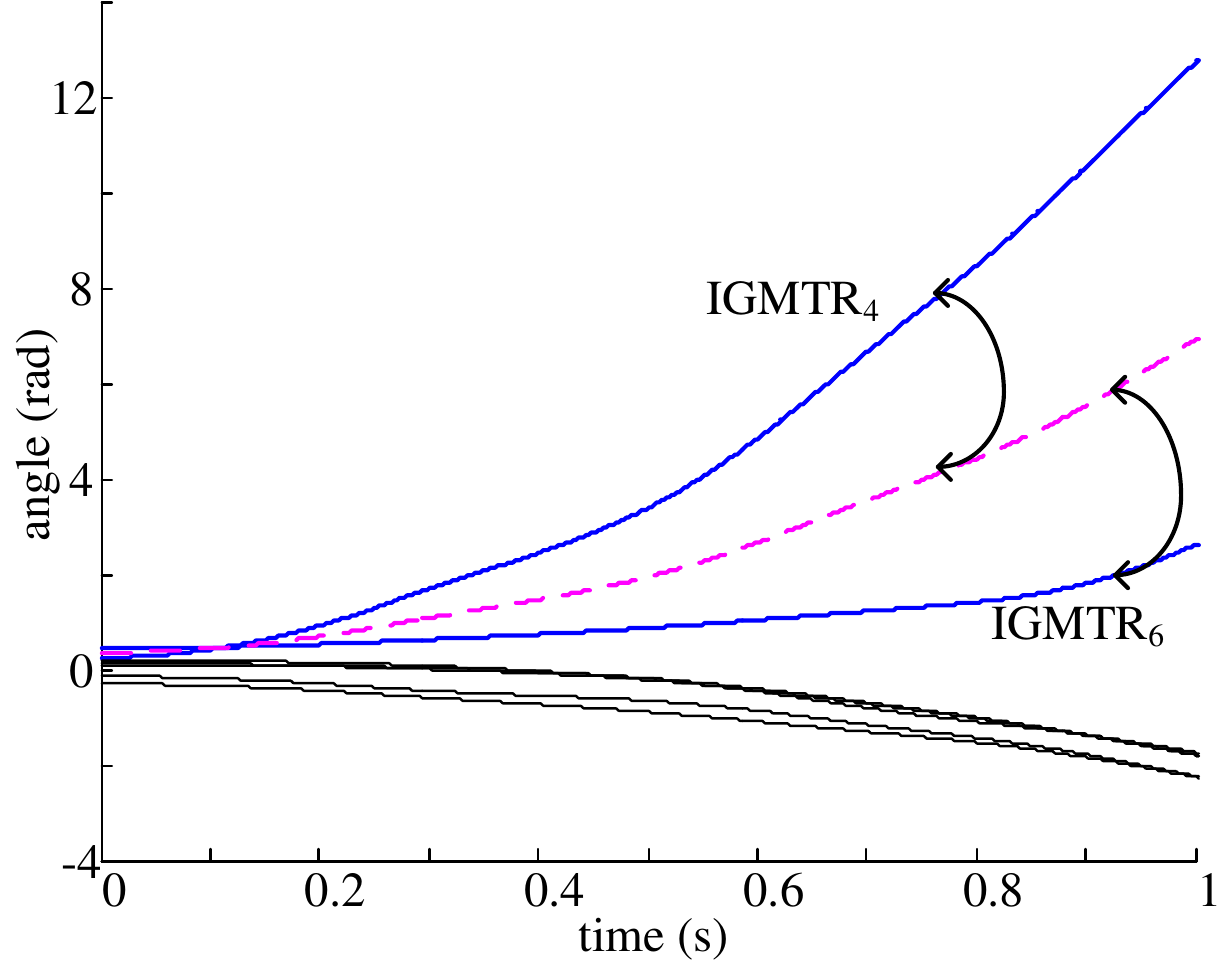}
  \caption{Demonstration of IGMTR [TS-6, bus-19, 0.260 s].} 
  \label{fig2}  
\end{figure}

 The IGMTR inside Group-CR is shown in Fig. \ref{fig2}. $\Omega_{\mathrm{CR}}$=\{Machine4, Machine 6\}. 
\vspace*{-0.5em}
\subsection{I-CR SYSTEM MODELING}  \label{section_IIC}
\begin{figure}[H]
  \centering
  \includegraphics[width=0.42\textwidth,center]{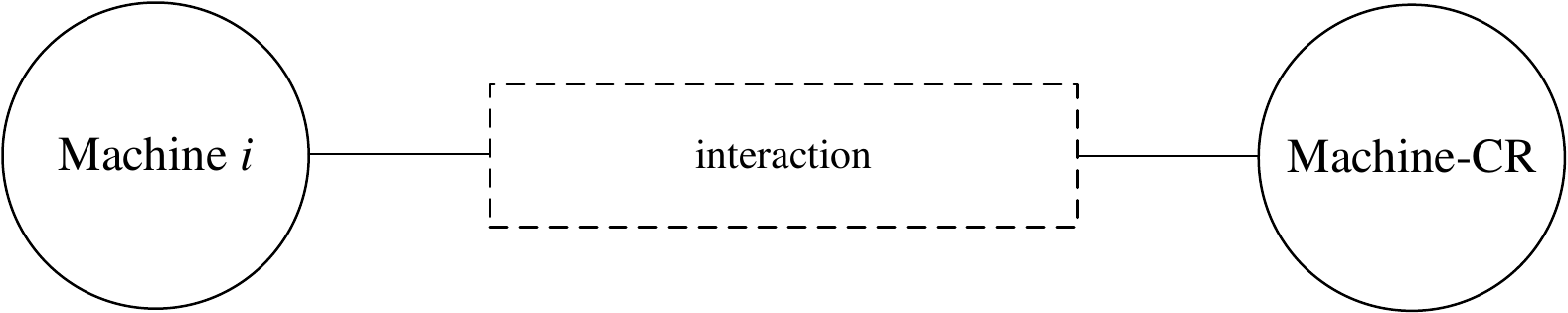}
  \caption{Formation of the I-CR system.} 
  \label{fig3}  
\end{figure}

Based on the inner-group machine monitoring, the variance of $\delta_{i\mbox{-}\mathrm{CR}}$ is modeled through the I-CR system. The formation of the I-CR system is shown in Fig. \ref{fig3}.
\vspace*{-0.5em}
Based on Eqs. (\ref{equ1}) and (\ref{equ2}), the relative motion between Machine \textit{i} and Machine-CR is depicted as
\begin{equation}
  \label{equ4}
  \left\{\begin{array}{l}
    \frac{d \delta_{i\mbox{-}\mathrm{CR}}}{d t}=\omega_{i\mbox{-}\mathrm{CR}} \\
    \\
    M_{i} \frac{d \omega_{i\mbox{-}\mathrm{CR}}}{d t}=f_{i\mbox{-}\mathrm{CR}}
    \end{array}\right.
\end{equation}
where
\begin{spacing}{1.5}
  \noindent$\omega_{i\mbox{-}\mathrm{CR}}=\omega_{i\mbox{-}\mathrm{SYS}}-\omega_{\mathrm{CR}\mbox{-}\mathrm{SYS}}$\\
  $f_{i\mbox{-}\mathrm{CR}}=f_{i\mbox{-}\mathrm{SYS}}-\frac{M_i}{M_{\mathrm{CR}}}f_{\mathrm{CR}\mbox{-}\mathrm{SYS}}$
\end{spacing}
Following the definition in Eq. (\ref{equ1}), in the COI-SYS reference, one can naturally obtain the following
\begin{equation}
  \label{equ5}
  \left\{\begin{array}{l}
    \sum_{i \in \Omega_{\mathrm{CR}}} M_{i} \delta_{i\mbox{-}\mathrm{CR}}=0 \\
    \\
    \sum_{i \in \Omega_{\mathrm{CR}}} M_{i} \omega_{i\mbox{-}\mathrm{CR}}=0 \\
    \\
    \sum_{i \in \Omega_{\mathrm{CR}}} f_{i\mbox{-}\mathrm{CR}}=0
    \end{array}\right.
\end{equation}
\par The inner-group-machine DLP (IGMDLP) is denoted as 
\begin{equation}
  \label{equ6}
  f_{i\mbox{-}\mathrm{CR}}=0
\end{equation}
\par In Eq. (\ref{equ6}), the IGMDLP of Machine \textit{i} depicts the point where the machine becomes unstable with respect to Machine-CR.

\subsection{INNER-GROUP MACHINE TRANSIENT ENERGY CONVERSION} \label{section_IID}
The IGMTE is defined in a typical Newtonian energy manner. The IGMTE of the inner-group machine is defined as
\begin{equation}
  \label{equ7}
  V_{i\mbox{-}\mathrm{CR}}=V_{KEi\mbox{-}\mathrm{CR}}+V_{PEi\mbox{-}\mathrm{CR}}
\end{equation}
where
\begin{spacing}{2}
    \noindent$V_{K E i\mbox{-}\mathrm{CR}}=\frac{1}{2} M_{i} \omega_{i\mbox{-}\mathrm{CR}}^{2}$\\
    $V_{P E i\mbox{-}\mathrm{CR}}=\int_{\delta_{i\mbox{-}\mathrm{CR}}^{s}}^{\delta_{i\mbox{-}\mathrm{CR}}}\left[-f_{i\mbox{-}\mathrm{CR}}^{(P F)}\right] d \delta_{i\mbox{-}\mathrm{CR}}$
\end{spacing}
In Eq. (\ref{equ7}), the conversion between IGMKE and IGMPE is used to measure the stability of the I-CR system.
\par The residual KE of the inner-group-machine \textit{i} at its corresponding MPP is denoted as
\begin{equation}
  \label{equ8}
  \begin{split}
    V_{K E i\mbox{-}\mathrm{CR}}^{R E}&=V_{K E i\mbox{-}\mathrm{CR}}^{c}-\Delta V_{P E i\mbox{-}\mathrm{CR}}\\
    &=A_{A C C i\mbox{-}\mathrm{CR}}-A_{D E C i\mbox{-}\mathrm{CR}}
  \end{split}
\end{equation}
where
\begin{spacing}{2}
  \noindent$V_{K E i\mbox{-}\mathrm{CR}}^{c}=\frac{1}{2} M_{i} \omega_{i\mbox{-}\mathrm{CR}}^{c 2}=A_{A C C i\mbox{-}\mathrm{CR}}$\\
  $\Delta V_{P E i\mbox{-}\mathrm{CR}}=\int_{\delta_{i\mbox{-}\mathrm{CR}}^{s}}^{\delta_{i\mbox{-}\mathrm{CR}}^{M  P P}}\left[-f_{i-\mathrm{CR}}^{(P F)}\right] d \delta_{i\mbox{-}\mathrm{CR}}-\\
  \int_{\delta_{i\mbox{-}\mathrm{CR}}^{s}}^{\delta_{i\mbox{-}\mathrm{CR}}^{c}}\left[-f_{i\mbox{-}\mathrm{CR}}^{(P F)}\right] d \delta_{i\mbox{-}\mathrm{CR}}=A_{DECi\mbox{-}\mathrm{CR}} $
\end{spacing}
\par In Eq. (\ref{equ8}), similar to the characteristics of the IMEAC and EMEAC, the inner-group-machine transient energy conversion is identical to the IGMEAC.
\par The stability characterizations of the inner-group machine are summarized as below.
\vspace*{0.5em}
\\
(i) From the perspective of transient energy conversion, the inner-group machine is evaluated to go unstable if the residual IGMKE occurs at its IGMPP.
\\ (ii) From the perspective of EAC, the inner-group machine is evaluated to go unstable if the acceleration area is larger than the deceleration area.
\par Detailed analysis about the IGMEAC will be given in Section \ref{section_III}.

\subsection{MECHANISMS OF THE INNER-GROUP MACHINE} \label{section_IIE}
From the analysis in Sections \ref{section_IIA} to \ref{section_IIC}, based on the transient stability paradigms, the IGMTR of each machine inside the group is modeled through the I-CR system. Against this background, the stability of inner-group machine can be characterized precisely through NEC.
\par
\begin{figure}[H]
  \centering
  \includegraphics[width=0.45\textwidth,center]{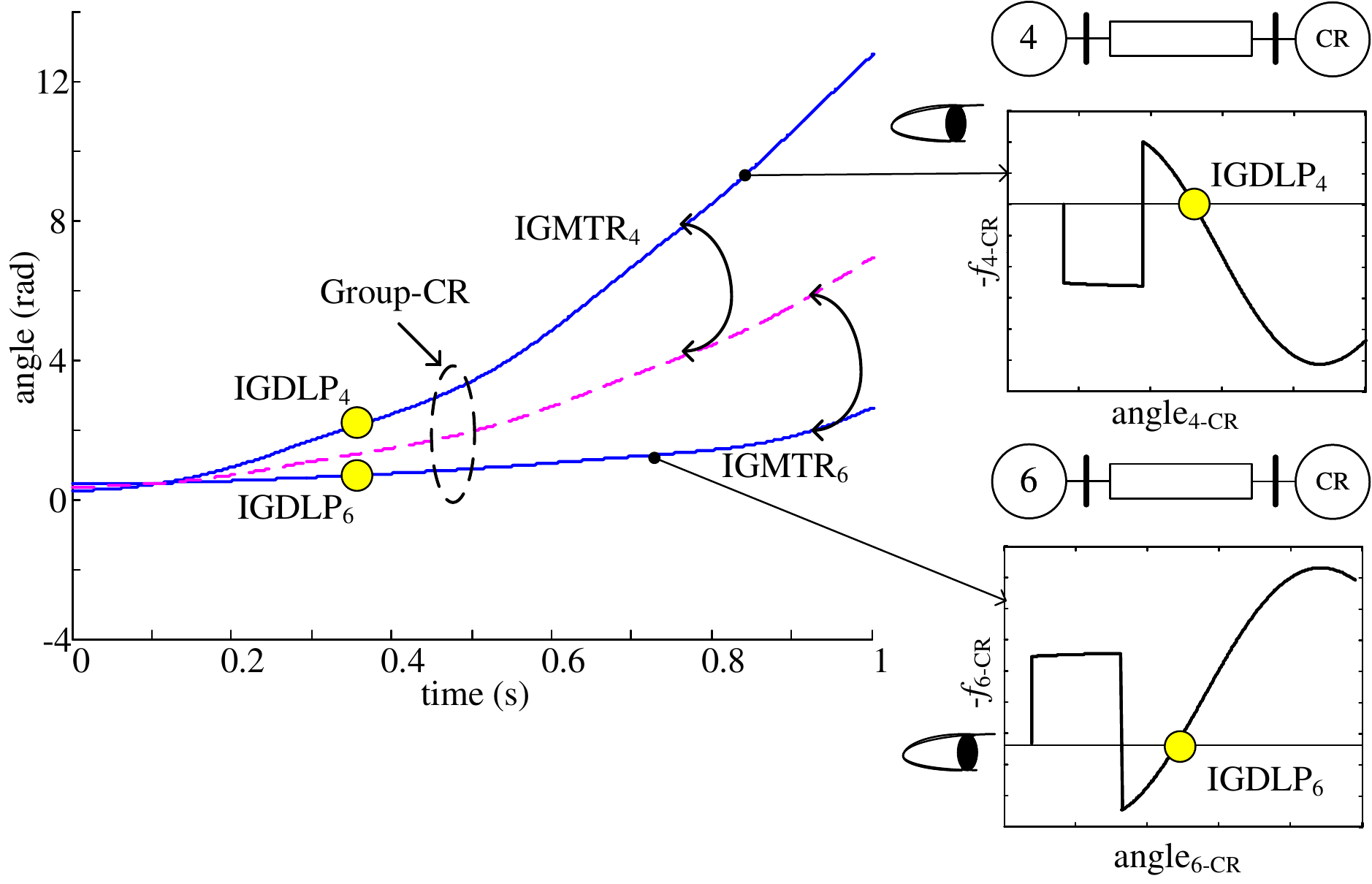}
  \caption{Mechanism of the inner-group machine [TS-6, bus-19, 0.260 s].} 
  \label{fig4}  
\end{figure}

 The mechanism of the inner-group machine is shown in Fig. \ref{fig4}.
\vspace*{-0.5em}
From Fig. \ref{fig4}, for the original system with \textit{n} machines, \textit{n} inner-group machines in the two groups can be formed. These inner-group machines physically do not exist in the system.
Instead, they are created from the differences between the original system and the equivalent system. Therefore, the stability of the inner-group machine can be used to measure the  ``level'' of the difference between the equivalent system and the original system. Based on this, one can naturally obtain the following
\par \textit{The equivalent system is close to the original system if all the inner-group machines are stable}.
\par \textit{The equivalent system is different from the original system if any inner-group machine becomes unstable}.
\vspace*{0.5em}
\par From the deductions above, it is clear that the transient behavior of the inner-group machine might have significant effect to the severity analysis in the equivalent-machine based TSA. Detailed analysis will be given in the following section.
\\ \\ \\
\section{TRANSIENT CHARACTERISTICS OF THE INNER-GROUP MACHINE } \label{section_III}
\subsection{STABLE INNER-GROUP MACHINE} \label{section_IIIA}
In most simulation cases, the equivalent system is close to the original system. Against this background, ``all'' the inner-group machine motions remain inseparable or even be twined in the COI-CR (and COI-NCR) reference. 
\par
\begin{figure}[H]
  \centering
  \includegraphics[width=0.4\textwidth,center]{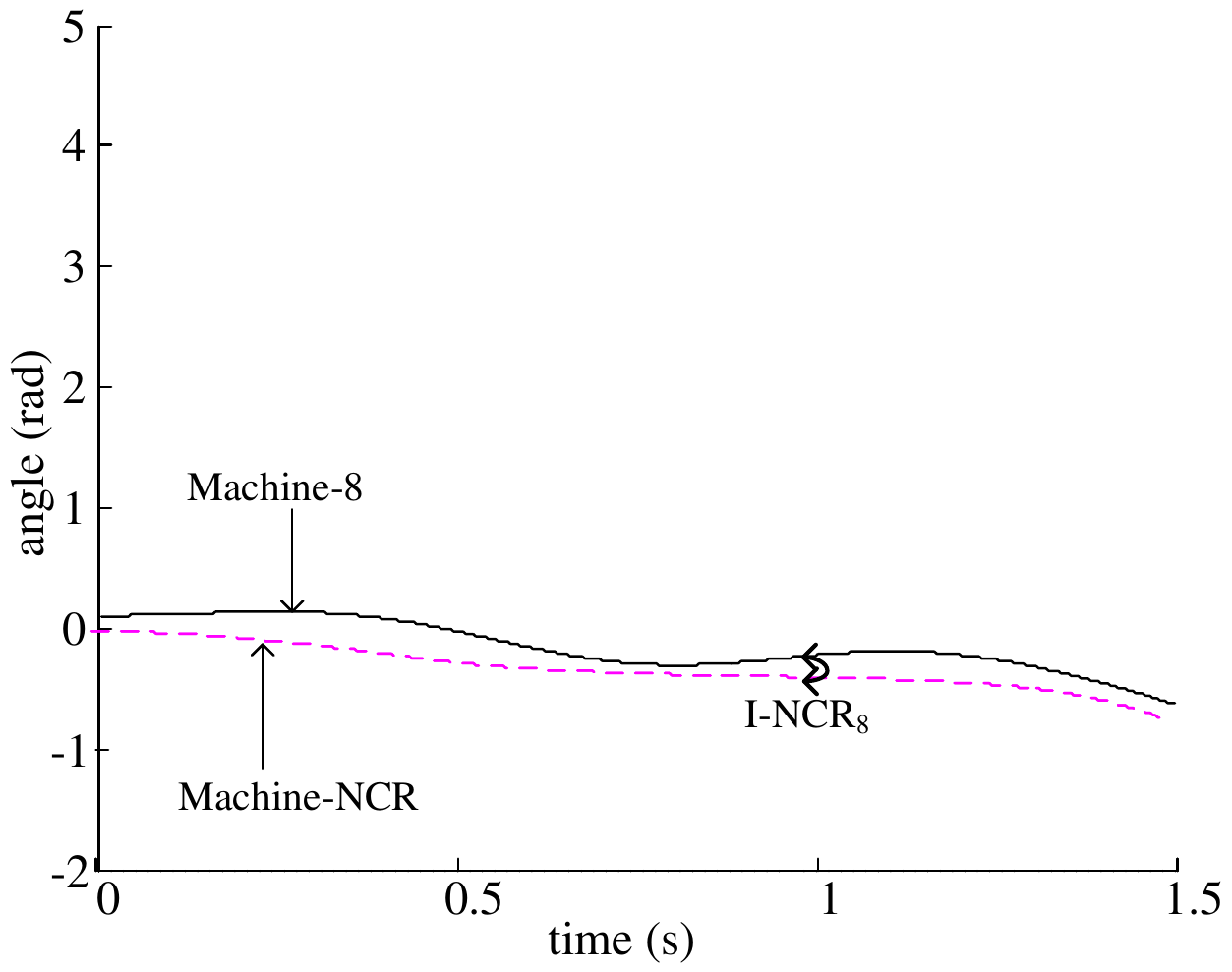}
  \caption{Inseparable inner-group motion [TS-1, bus-4, 0.447 s].} 
  \label{fig5}  
\end{figure}

\begin{figure}[H]
  \centering
  \includegraphics[width=0.38\textwidth,center]{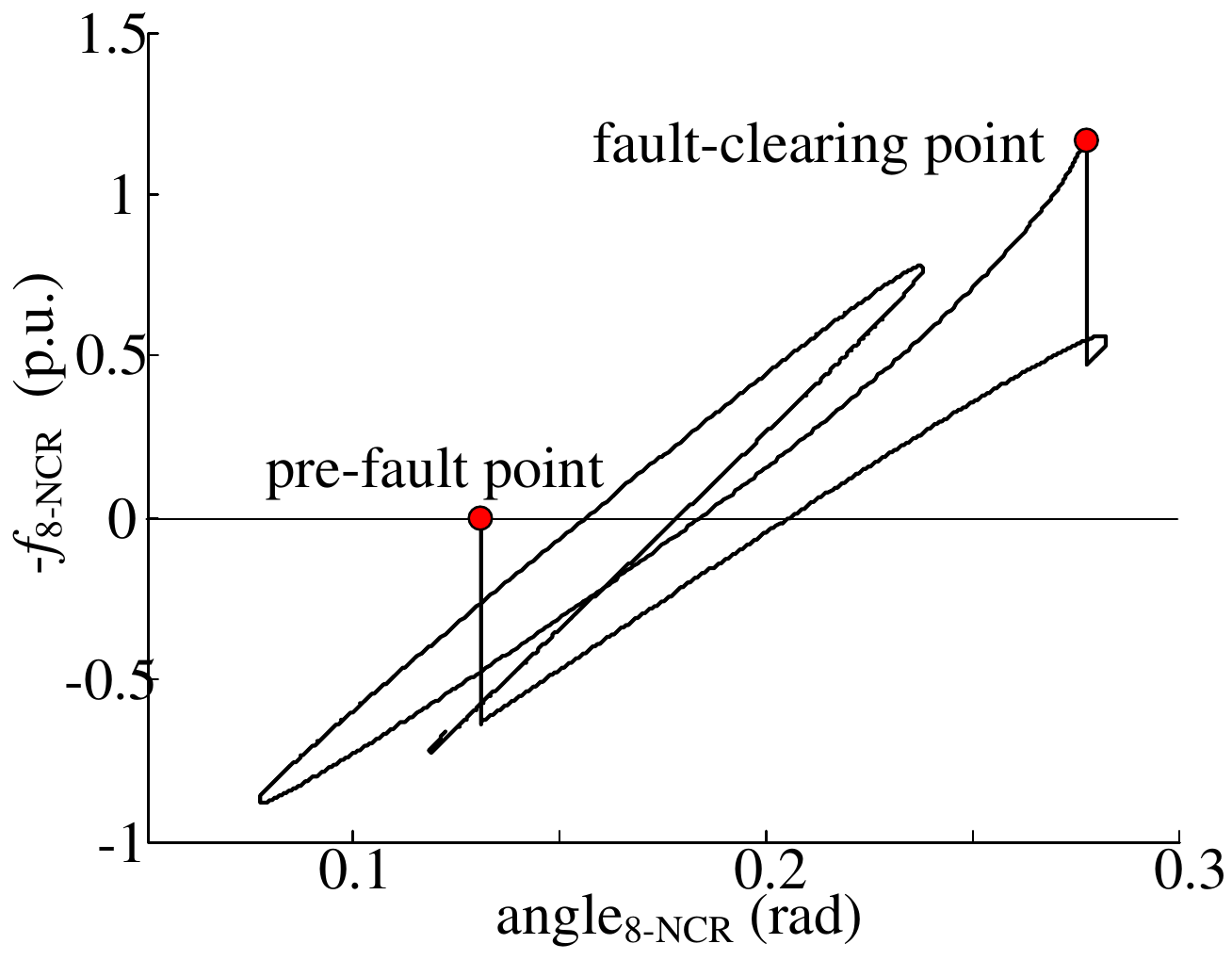}
  \caption{Kimbark curve of $\text{I-NCR}_8$ system [TS-1, bus-4, 0.447s].} 
  \label{fig6}  
\end{figure}

 Demonstration about the inseparable inner-group motion of the $\text{I-NCR}_8$ system (formed by Machine 8 and Machine-NCR) is shown in Fig. \ref{fig5}.
The original system trajectory and the equivalent system trajectory in the COI-SYS reference is given in Refs. \cite{2} and \cite{3}, respectively.
$\Omega_{\mathrm{CR}}$=\{Machine 2, Machine 3\}. The Kimbark curve of the $\text{I-NCR}_8$ system is shown in Fig. \ref{fig6}.
\vspace*{-1em}
\vspace*{-0.5em}
From Fig. \ref{fig5}, in this case Machine 8 shows very slight relative motion with respect to Machine-NCR.
Therefore, Machine 8 can be seen as a ``non-critical'' machine in the ``COI-NCR'' reference.
The Kimbark curve of the $\text{I-NCR}_8$ system can neither show ``acceleration-deceleration'' characteristic. The inner-group machine remains stable.

\subsection{UNSTABLE INNER-GROUP MACHINE} \label{section_IIIB}
Under certain distinctive simulation environments, the equivalent system might be significantly different from the original system. Against this background, the inner-group machine motion might become separable, and the inner-group machine instability will occur.
\par
\begin{figure}[H]
  \centering
  \includegraphics[width=0.38\textwidth,center]{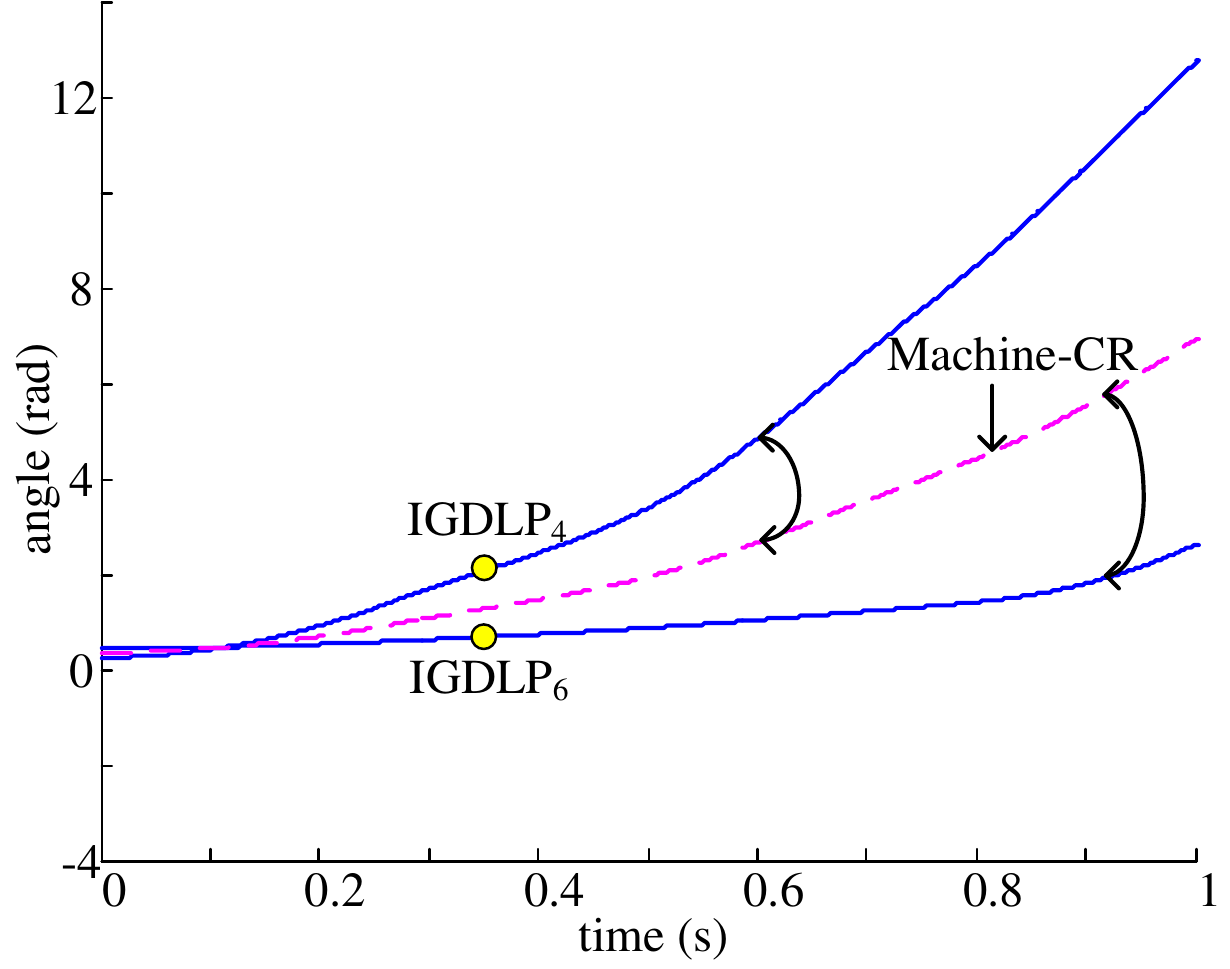}
  \caption{Separable inner-group motion [TS-6, bus-19, 0.260 s].} 
  \label{fig7}  
\end{figure}

 Demonstration about the separable inner-group motion is shown in Fig. \ref{fig7}. The fault is set as [TS-6, bus-19, 0.260 s]. $\Omega_{\mathrm{CR}}$ is \{Machine 4, Machine 6\}.
The stability analysis of Machine-CR, and Machines 4 and 6 in the COI-SYS reference was already given in Ref. \cite{5}.
The Kimbark curves of $\text{I-CR}_4$ and $\text{I-CR}_6$ are shown in Figs. \ref{fig8} (a) and (b), respectively. 
\vspace*{-1em}
\begin{figure} [H]
  \centering 
  \subfigure[]{%
  \label{fig8a}
    \includegraphics[width=0.37\textwidth]{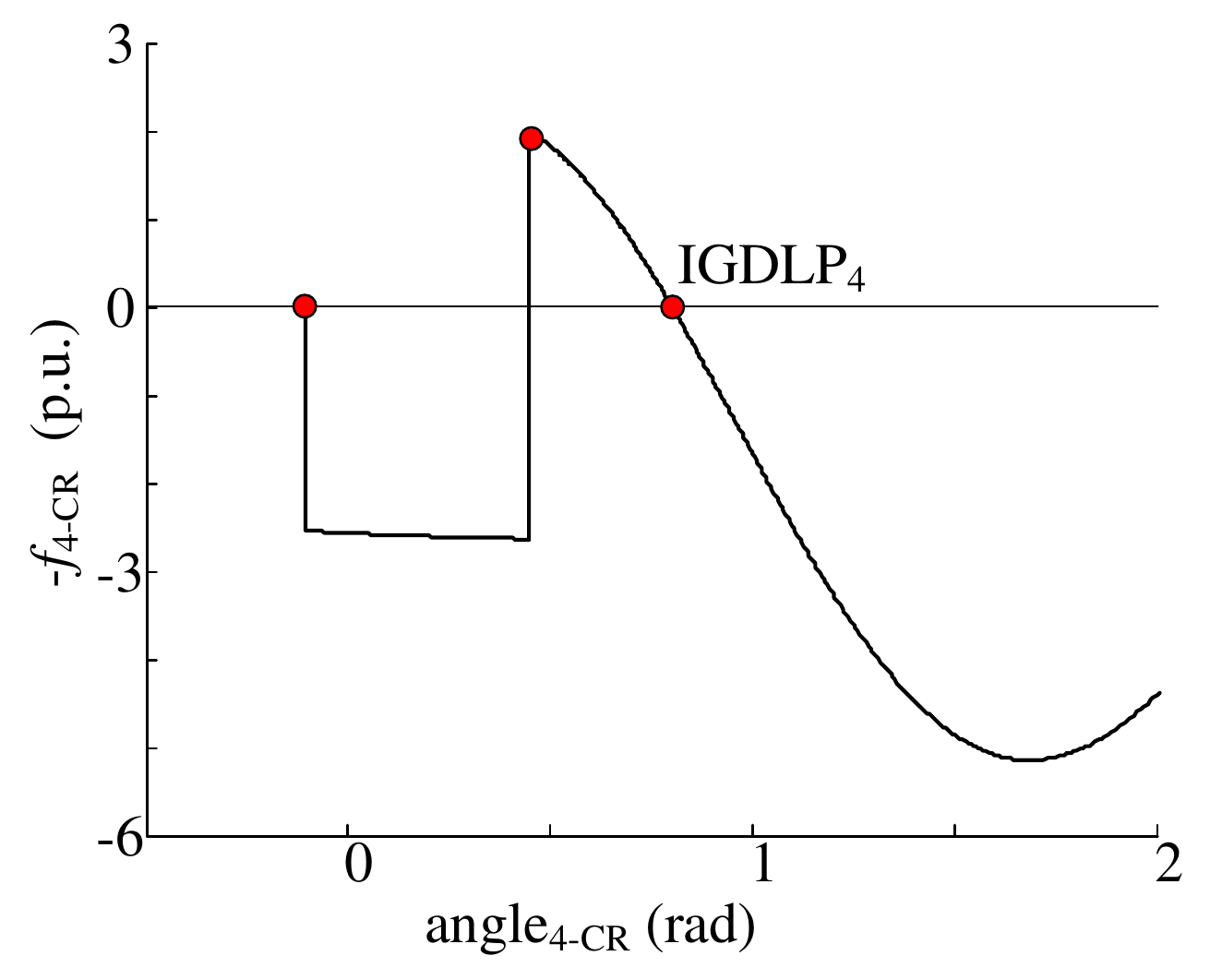}}%
\end{figure} 
\addtocounter{figure}{-1}
\vspace*{-2em}       
\begin{figure} [H]
  \addtocounter{figure}{1}      
  \centering 
  \subfigure[]{%
    \label{fig8b}
    \includegraphics[width=0.37\textwidth]{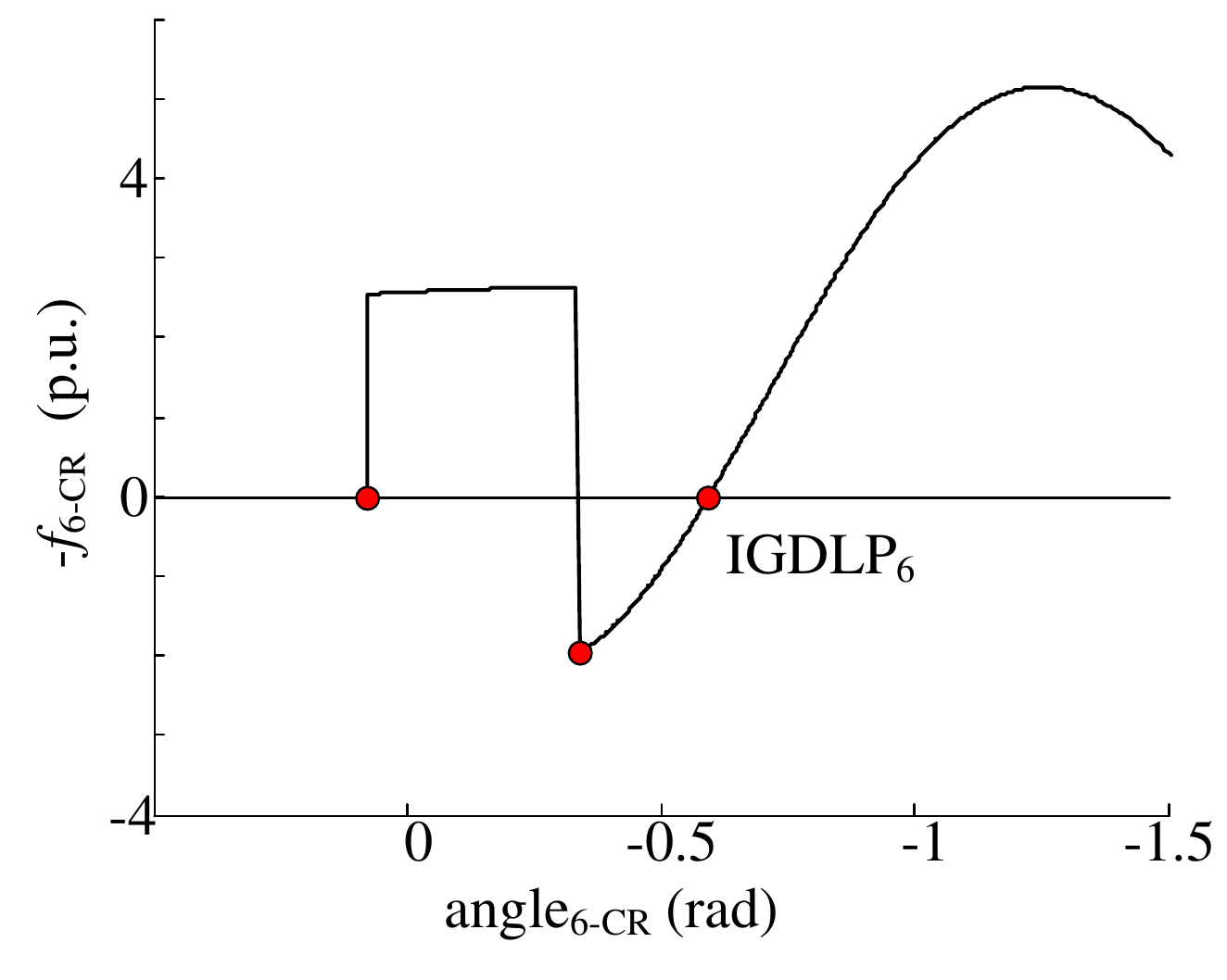}}%
  \caption{Kimbark curve of the I-CR system [TS-6, bus-19, 0.260 s]. (a) $\text{I-CR}_4$. (b) $\text{I-CR}_6$}%
  \label{fig8}
\end{figure}
\vspace*{-0.5em}
From Fig. \ref{fig7}, in this case Machines 4 and 6 show fierce relative motions with respect to Machine-CR.
Therefore, the two machines can be seen as two ``critical'' machines in the ``COI-CR'' reference”. The Kimbark curves of the two machines also show clear ``acceleration-deceleration'' characteristics. The two inner-group machines become unstable.
\par The two machines becoming unstable inside Group-CR fully indicates that Machine-CR cannot fully represent the original system. Against this background, if the system operator only monitors the transient behavior of Machine-CR, the important inner-group instabilities will be neglected. This is certain to cause significant effect to the severity evaluation of the original system if it is simply replaced with the equivalent system.

\subsection{OCCURRENCE OF IGMDLP} \label{section_IIIC}
For the machines inside Group-CR, the occurrence of IGMLP is denoted as
\begin{equation}
  \label{equ9}
  f_{i\mbox{-}\mathrm{CR}}=f_{i\mbox{-}\mathrm{SYS}}-\frac{M_{i}}{M_{\mathrm{CR}}} f_{\mathrm{CR}\mbox{-}\mathrm{SYS}}=0
\end{equation}
\par The motion of an individual machine in the COI-CR reference is more complicated than that in the COI-SYS reference.
Taking the case in Fig. \ref{fig7} as an example, it can be found that the two machines and their equivalent Machine-CR become separable in the COI-SYS reference. Based on this, it is certain that the separation of the machine in the COI-CR reference will be ``slighter'' than that in the COI-SYS reference. Generally, the following holds for the IGMDLP
\vspace*{0.5em}
\par \textit{The IGMDLP of the machine occurs later than EDLP}.
\par \textit{The IGMDLP of the machine also occurs later than the IDLP of the machine}.
\vspace*{0.5em}
\par In fact, the deduction above is also the reflection of the ``slighter'' separation of the inner-group machine in the COI-CR reference. The clarifications are given as below.
\\ \textit{Clarification}: Based on Eq. (\ref{equ9}), for the inner-group motions inside $\Omega_{\mathrm{CR}}$, the primary condition for $f_{k\mbox{-}\mathrm{CR}}$ becoming zero can be denoted as
\begin{equation}
  \label{equ10}
  f_{k\mbox{-}\mathrm{SYS}}f_{\mathrm{CR}\mbox{-}\mathrm{SYS}}>0
\end{equation}
\par Generally, Eq. (\ref{equ10}) can be satisfied only when both Machine \textit{k} and Machine-CR become unstable in the COI-SYS reference ($f_{k\mbox{-}\mathrm{SYS}}>0$, and $f_{\mathrm{CR}\mbox{-}\mathrm{SYS}}>0$).
This reveals that Machine \textit{k} and Machine-CR already become unstable for a while in the COI-SYS reference when the $\text{IGMDLP}_k$ occurs. Thus the deduction is clarified.
\par
\begin{figure}[H]
  \centering
  \includegraphics[width=0.38\textwidth,center]{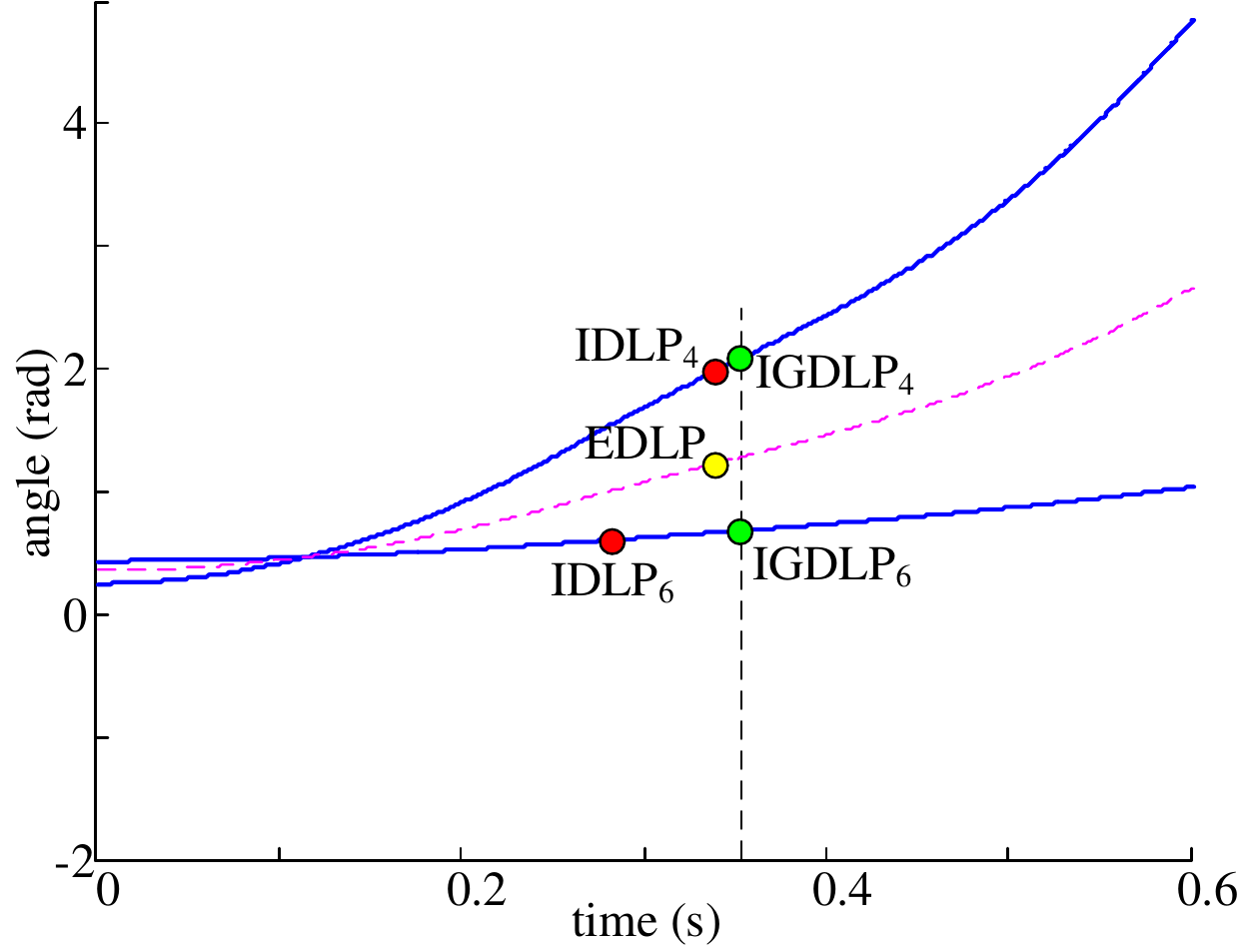}
  \caption{The occurrence of IGMDLP [TS-6, bus-19, 0.260 s].} 
  \label{fig9}  
\end{figure}

 The occurrence of IGMDLP is shown in Fig. \ref{fig9}. Note that $\text{IDLP}_6$, $\text{IDLP}_4$ and $\text{EDLP}_{\text{CR}}$ are defined in the COI-SYS reference, while $\text{IGMDLP}_6$ and $\text{IGMDLP}_4$ are defined in the COI-CR reference.
The $f_i$ of Machines 4 and 6 at different time points are shown in Table \ref{table1}. The dominant pattern is \{Machine 4, Machine 6\}.
\vspace*{-1em}
\begin{table}[H]
  \centering
  \small
  \caption{ $f_i$ of machines inside Group-CR}
  \begin{tabular}{cccc}
  \hline
  \begin{tabular}[c]{@{}c@{}}Machine \\ NO.\end{tabular} & \begin{tabular}[c]{@{}c@{}}\textit{f} at\\ $\text{IDLP}_6$ (p.u.)\end{tabular} & \begin{tabular}[c]{@{}c@{}}\textit{f} at\\ $\text{EDLP}_{\text{CR}}$ (p.u.)\end{tabular} & \begin{tabular}[c]{@{}c@{}}\textit{f} at\\ $\text{IDLP}_4$ (p.u.)\end{tabular} \\ \hline
  Machine 6                                              & {\ul 0.0000}                                                       & -0.2742                                                           & 0.2661                                                            \\
  Machine-CR                                             & -2.9310                                                             & {\ul 0.0000}                                                      & 0.2661                                                            \\
  Machine 4                                              & -2.9310                                                             & 0.2742                                                            & {\ul 0.0000}                                                        \\ \hline
  \end{tabular}
  \label{table1}
\end{table}
\vspace*{-0.5em}
\noindent 1) THE OCCURRENCE OF IGMDLP
\par Because $\Omega_{\mathrm{CR}}$ is formed by Machines 4 and 6, following Eq. (\ref{equ5}), one can obtain that
\begin{equation}
  \label{equ11}
  f_{4\mbox{-}\mathrm{CR}}+f_{6\mbox{-}\mathrm{CR}}=0
\end{equation}
\par From Eq. (\ref{equ11}), both $f_{4\mbox{-}\mathrm{CR}}$ and $f_{6\mbox{-}\mathrm{CR}}$ will reach zero simultaneously.
This indicates that both $\text{IGMDLP}_4$ and $\text{IGMDLP}_6$ occur simultaneously along time horizon.
\par
\begin{figure}[H]
  \centering
  \includegraphics[width=0.42\textwidth,center]{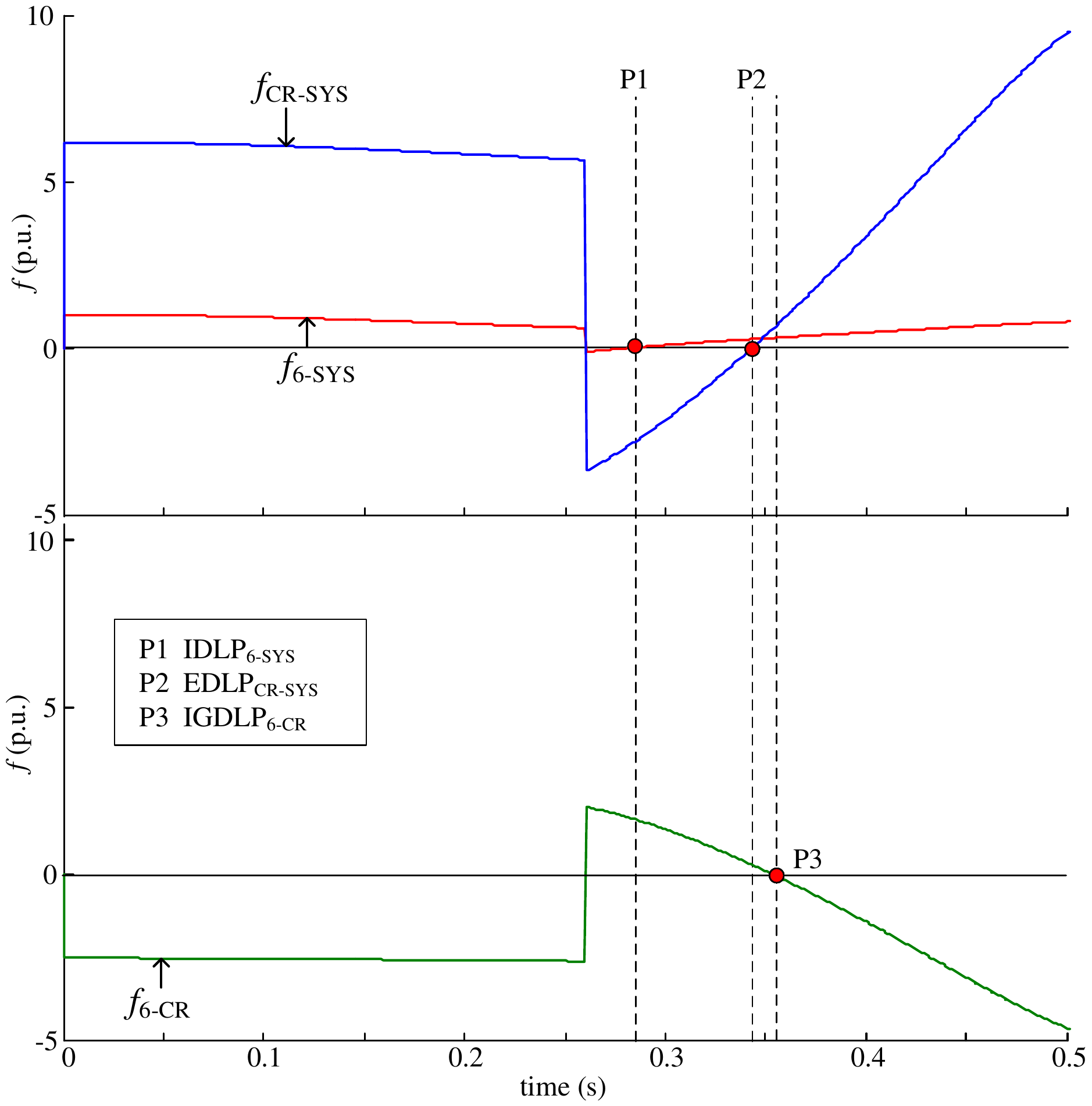}
  \caption{Variance of \textit{f} along time horizon [TS-6, bus-19, 0.260s].} 
  \label{fig10}  
\end{figure}

 Taking the occurrence of $\text{IGMDLP}_6$ as an example, the variance of $f_{6\mbox{-}\mathrm{SYS}}$ and $f_{\mathrm{CR}\mbox{-}\mathrm{SYS}}$ along time horizon is shown in Fig. \ref{fig10}.
Note that these curves are not the Kimbark curves because they are depicted in the \textit{t-f} space.
\vspace*{-0.5em}
From Fig. \ref{fig10}, the $\text{IGMDLP}_6$ occurs only when both $f_{6\mbox{-}\mathrm{SYS}}$ and $f_{\mathrm{CR}\mbox{-}\mathrm{SYS}}$ are positive.
This also indicates that Machine 6 and Machine-CR already become unstable in the COI-SYS reference for a while when IGMDLP occurs. This fully reflects that the separation of Machine 6 in the COI-CR reference is ``slighter'' than that in the COI-SYS reference. The analysis above can be simply extended to the case of $\text{IGMDLP}_4$.
\par  The inner-group machine motion as analyzed in this section is quite ideal because it is first-swing unstable. In fact, the inner-group motion might also become a complicated multi-swing instability problem. This will be analyzed in the case study.

\section{CASE STUDY} \label{section_IV}
\subsection{TEST BED} \label{section_IVA}
\begin{figure}[H]
  \centering
  \includegraphics[width=0.42\textwidth,center]{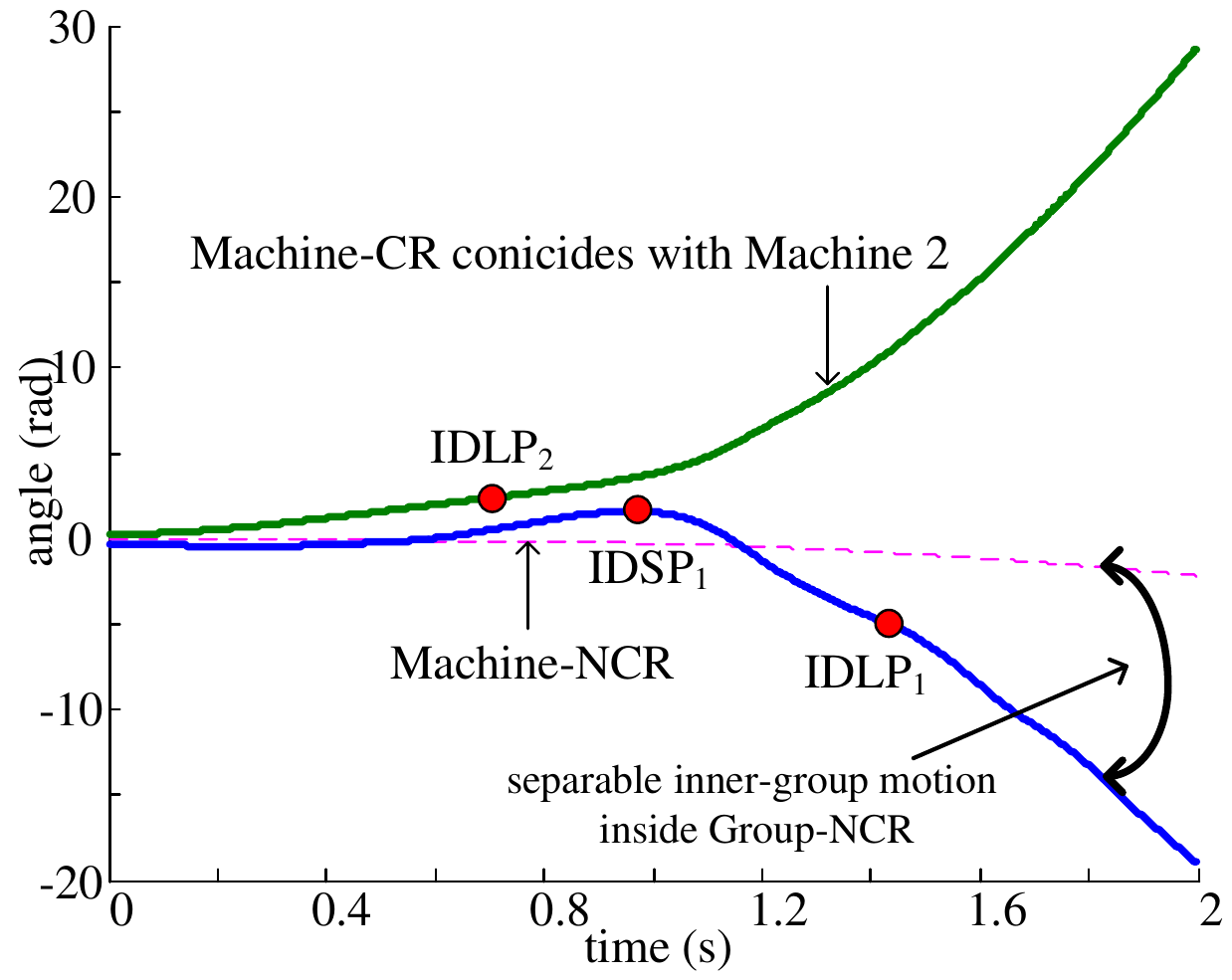}
  \caption{Multi-swing inner-group instability in Group-NCR [TS-2, bus-12, 0.550 s].} 
  \label{fig11}  
\end{figure}

A complicated system trajectory [TS-2, bus-12, 0.550 s] is given to demonstrate the multi-swing instability of the inner-group machine. The individual-machine based transient stability analysis was already shown in Ref. \cite{2}. In this section only Machines 1 and 2 are shown in the figure for clearance, as in Fig. \ref{fig11}.
\vspace*{-1.5em}
\subsection{EQUIVLAENT SYSTEM STABILITY} \label{section_IVB}
From Fig. \ref{fig11}, it is quite clear that the original system trajectory separates complicatedly. The two most possible group separation patterns are given as follows
\vspace*{0.5em}
\\ Pattern-1: $\Omega_{\mathrm{CR}}$=\{Machine 2\}, $\Omega_{\mathrm{NCR}}$=\{rest\};
\\ Pattern-2: $\Omega_{\mathrm{NCR}}$=\{Machine 1\}, $\Omega_{\mathrm{CR}}$=\{rest\};
\vspace*{0.5em}
\par The equivalent system in the two patterns are shown in Figs.\ref{fig12} (a) and (b), respectively.
\begin{figure} [H]
  \centering 
  \subfigure[]{%
  \label{fig12a}
    \includegraphics[width=0.37\textwidth]{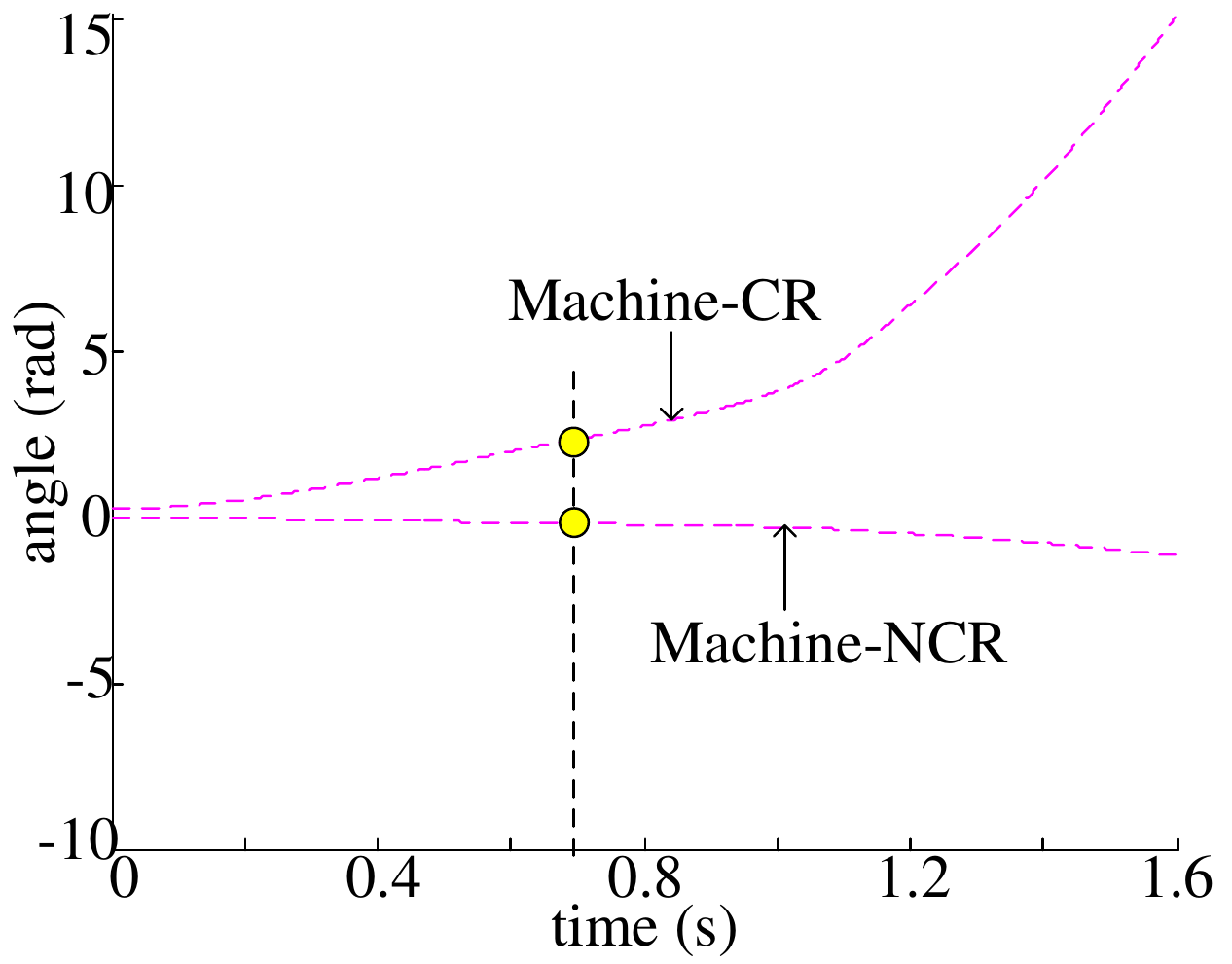}}%
\end{figure} 
\addtocounter{figure}{-1}
\vspace*{-2em}       
\begin{figure} [H]
  \addtocounter{figure}{1}      
  \centering 
  \subfigure[]{%
    \label{fig12b}
    \includegraphics[width=0.37\textwidth]{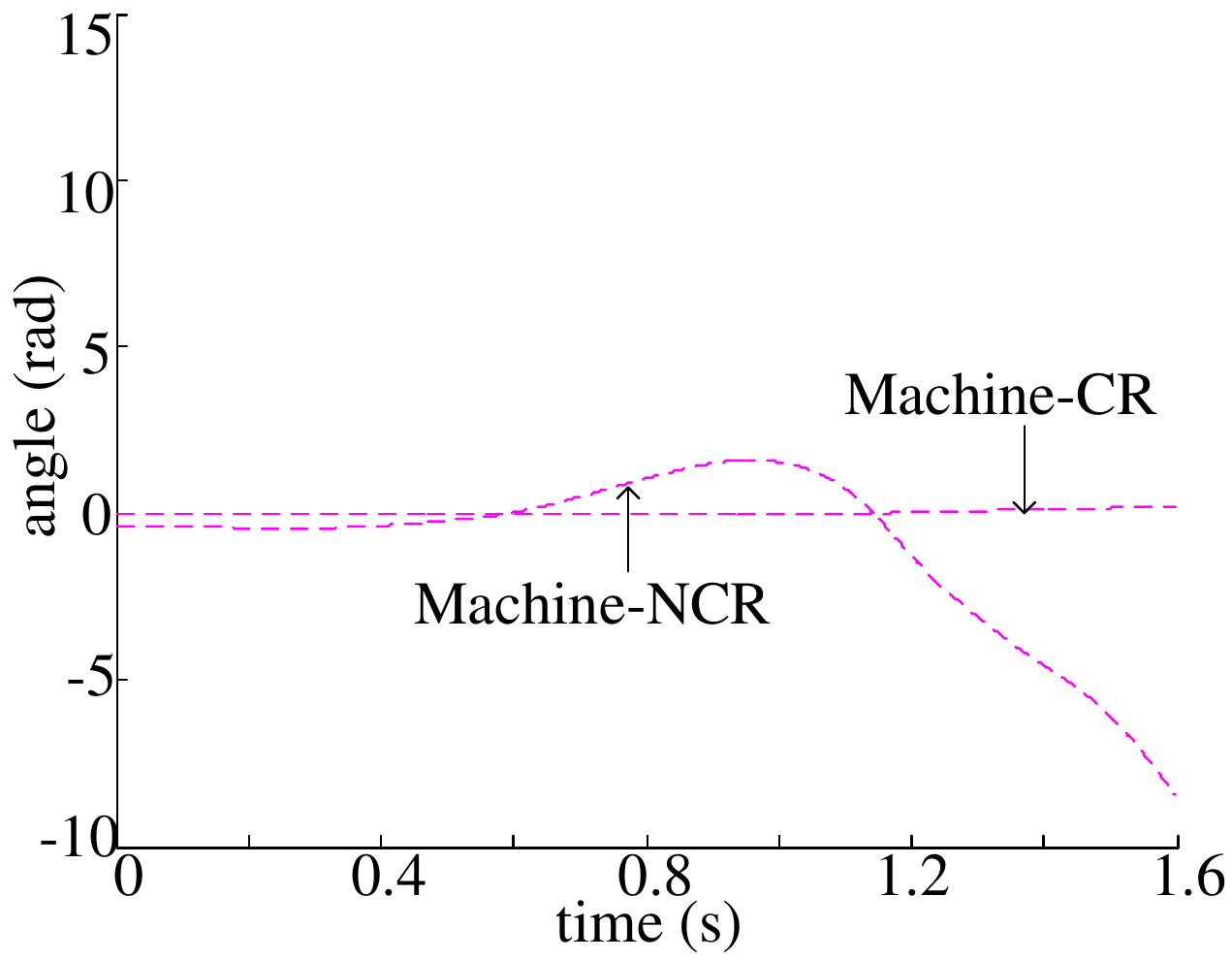}}%
  \caption{Equivalent system [TS-2, bus-12, 0.550 s]. (a) Pattern-1. (b) Pattern-2.}%
  \label{fig12}
\end{figure}
\vspace*{-0.5em}
Through the computation of $\eta_{\mathrm{CR}}$, Pattern-1 is finally identified as the dominant pattern. Pattern-1 is also shown in Fig. \ref{fig11}. The equivalent-machine based TSA is given as
\vspace*{0.5em}
\\ \textit{$\text{EDLP}_{\text{CR}}$ occurs (0.673 s)}: Machine-CR goes unstable, and thus the equivalent system is evaluated to go unstable.
\vspace*{0.5em}
\par In this case Group-CR is formed by only Machine 2, and thus $\text{EDLP}_{\text{CR}}$ coincides with $\text{IDLP}_2$. In addition, based on the mirror system \cite{3}.
\par From Fig. \ref{fig11}, the differences between the original system and the equivalent system are quite large, especially inside Group-NCR. This fully indicates that the inner-group motion inside Group-NCR is fierce. The inner-group machine instabilities will occur inside Group-NCR.

\subsection{INNER-GROUP MACHINE STABILITY} \label{section_IVC}
Following the definitions of the inner-group machine motion as analyzed in Section \ref{section_II}, nineteen inner-group machines can be found inside Group-NCR. From Fig. \ref{fig11}, Machine 1 separates from Machine-NCR as the relative motion between the Machine 1 and Machine-NCR reaches -20.1 rad at 2.000 s, which indicates that the inner-group instability occurs in the I-NCR system.
The rest of the eighteen inner-group motions inside $\Omega_{\mathrm{NCR}}$ maintain inseparable.
\par
\begin{figure}[H]
  \centering
  \includegraphics[width=0.38\textwidth,center]{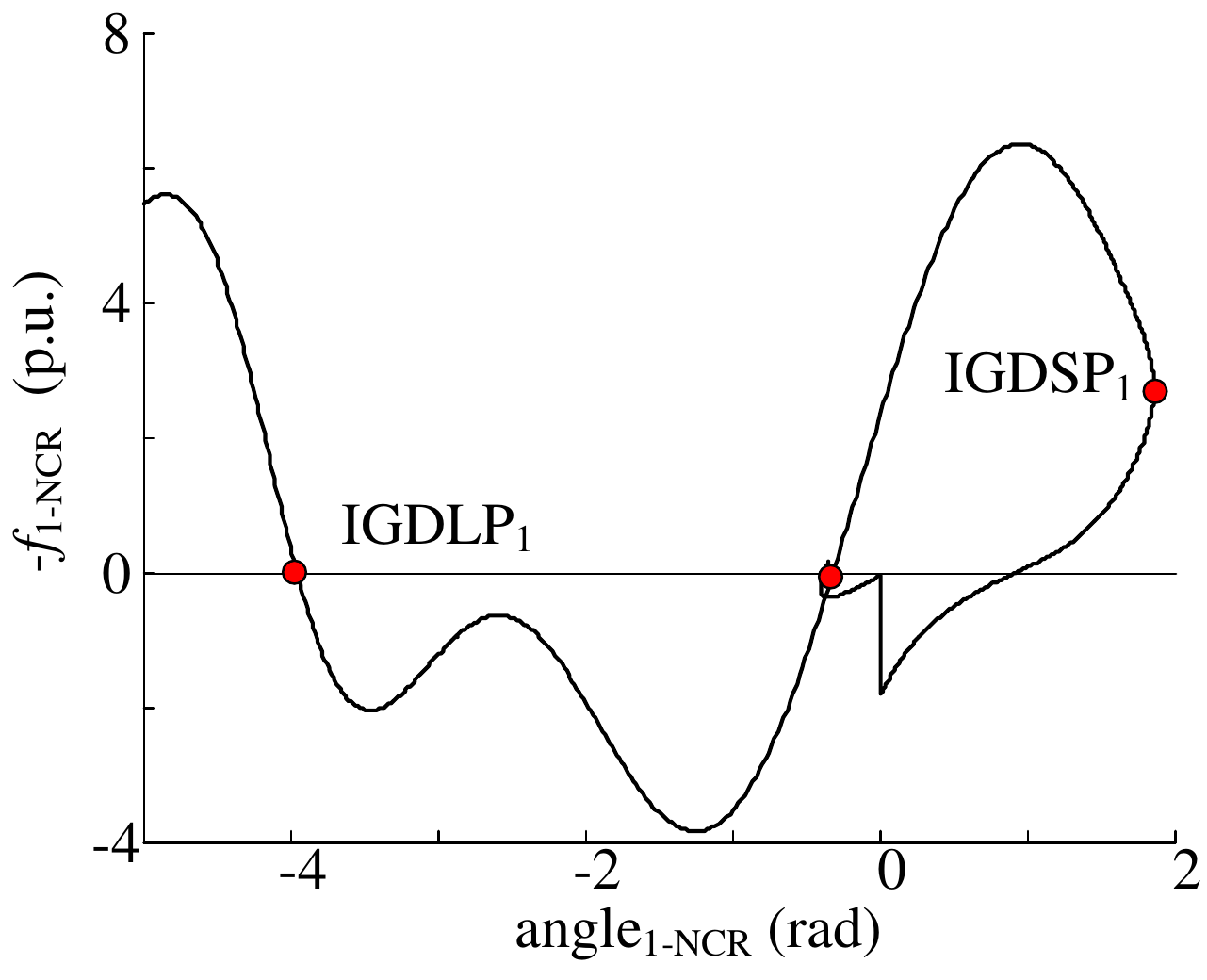}
  \caption{Kimbark curve of the $\text{I-NCR}_1$ system [TS-2, bus-12, 0.550s].} 
  \label{fig13}  
\end{figure}

\begin{figure}[H]
  \centering
  \includegraphics[width=0.38\textwidth,center]{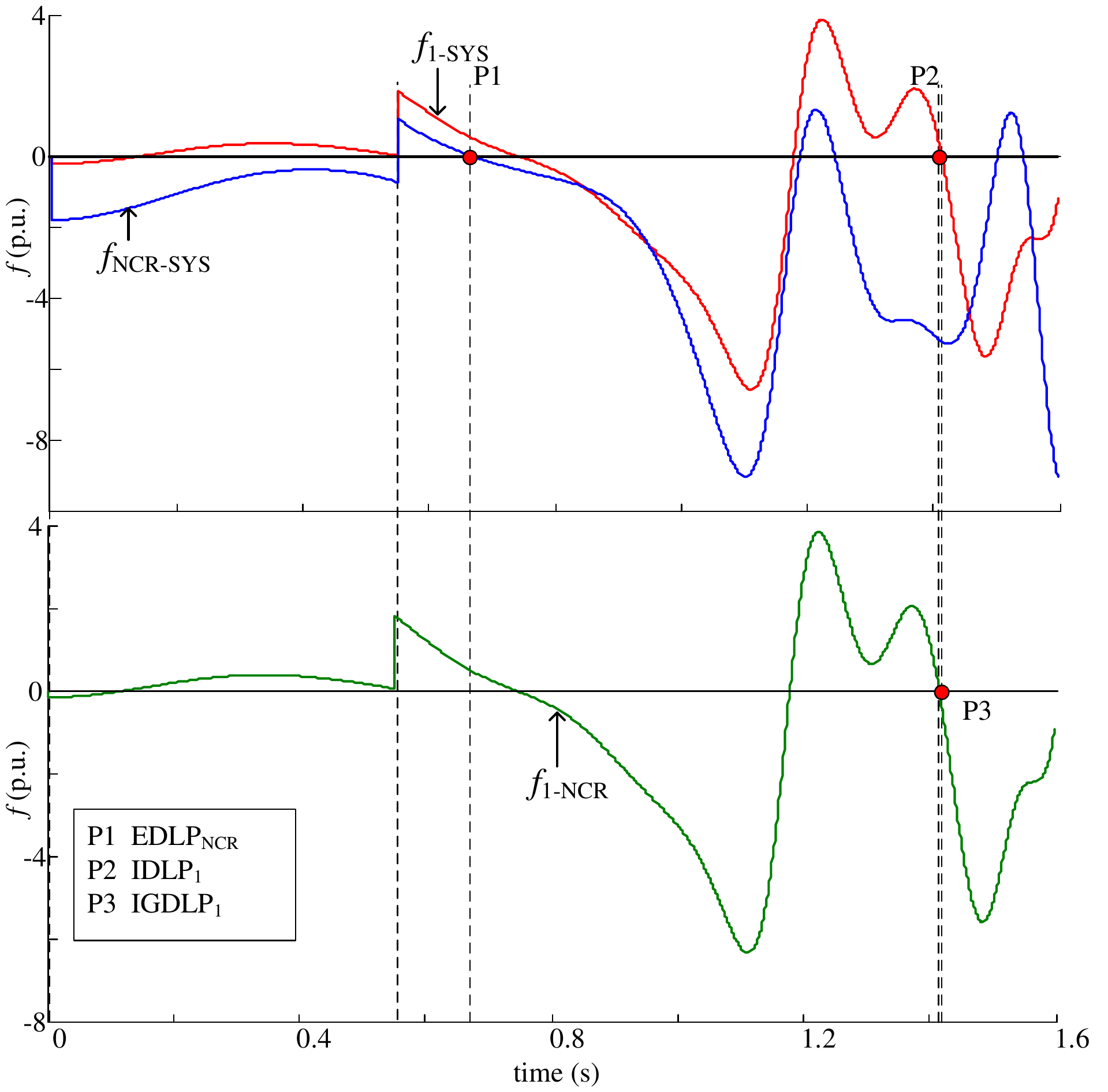}
  \caption{Variance of \textit{f} along time horizon.} 
  \label{fig14}  
\end{figure}

 Following the analysis in Section \ref{section_II}, based on the transient stability paradigms, this IGMTR variance is modeled through the $\text{I-NCR}_1$ system. The Kimbark curve of $\text{I-NCR}_1$ is shown in Fig. \ref{fig13}. The occurrence of the $\text{IGMDLP}_1$ is shown in Fig. \ref{fig14}.
\vspace*{-1em}
\vspace*{-0.5em}
From Figs. \ref{fig13} and \ref{fig14}, it is clear that multi-swing instability occurs in the $\text{I-NCR}_1$ system. At the moment that $\text{IGMDLP}_{\text{1-NCR}}$ occurs, the following holds
\begin{equation}
  \label{equ12}
  f_{1\mbox{-}\mathrm{NCR}}=f_{1\mbox{-}\mathrm{SYS}}-\frac{M_{1}}{M_{\mathrm{NCR}}} f_{\mathrm{NCR}\mbox{-}\mathrm{SYS}}=0
\end{equation}
\par From Eq. (\ref{equ12}) and Fig. \ref{fig14}, because $M_1/M_{\mathrm{NCR}}$ is quite small in this case (0.0275), $f_{1\mbox{-}\mathrm{SYS}}$ and $f_{1\mbox{-}\mathrm{NCR}}$ become quite close along time horizon.
This also indicates that $\text{IGMDLP}_1$ (1.1416 s) and $\text{IDLP}_1$ (1.1414 s) occur quite close along time horizon.
\par  Based on the analysis above, in this distinctive simulation case, Machine 1 becomes inner-group unstable inside Group-NCR. Against this background, the severity of the original system cannot be simply replaced with the equivalent system. Comparatively, the individual-machine based TSA without any equivalence will become more precise and flexible when inner-group machine instability occurs.

\section{COMPANION BETWEEN INDIVIDUAL MACHINE AND EQUIVALENT MACHINE IN ACTUAL TSA ENVIRONMENT} \label{section_V}
\subsection{``COMPROMISE'' IN THE EQUIVALENT MACHINE} \label{section_VA}
Compared with the theoretical analysis of the inner-group machine, i.e., the difference between the original system and equivalent system, analyzing the relationship between the individual-machine and the equivalent machine in actual TSA environment is more ``technical''.
\par Following the analysis in Refs. \cite{2} and \cite{3}, The stability evaluations of the original system and the equivalent system are based on the individual-machine and the equivalent machine, respectively. Both the individual-machine and the equivalent machine strictly follow the machine paradigms. These strict followings of the paradigms also bring the both the stability-characterization advantage and the trajectory-depiction advantage in TSA. However, note that the advantages of the individual-machine and the equivalent machine are based on the original system and the equivalent system, respectively.
\par The analysis above emerges the following question
\vspace*{0.5em}
\par \textit{What is the relationship between the individual-machine and the equivalent machine that both show advantages in TSA}?
\vspace*{0.5em}
\par Following the original definition of the equivalent machine that the machine is the ``motion equivalence'' of all the individual machines inside each group, it is clear that the equivalent machine will show the ``compromise'' characteristics between the stability and severity of the system compared with individual machine.
\par
\begin{figure}[H]
  \centering
  \includegraphics[width=0.42\textwidth,center]{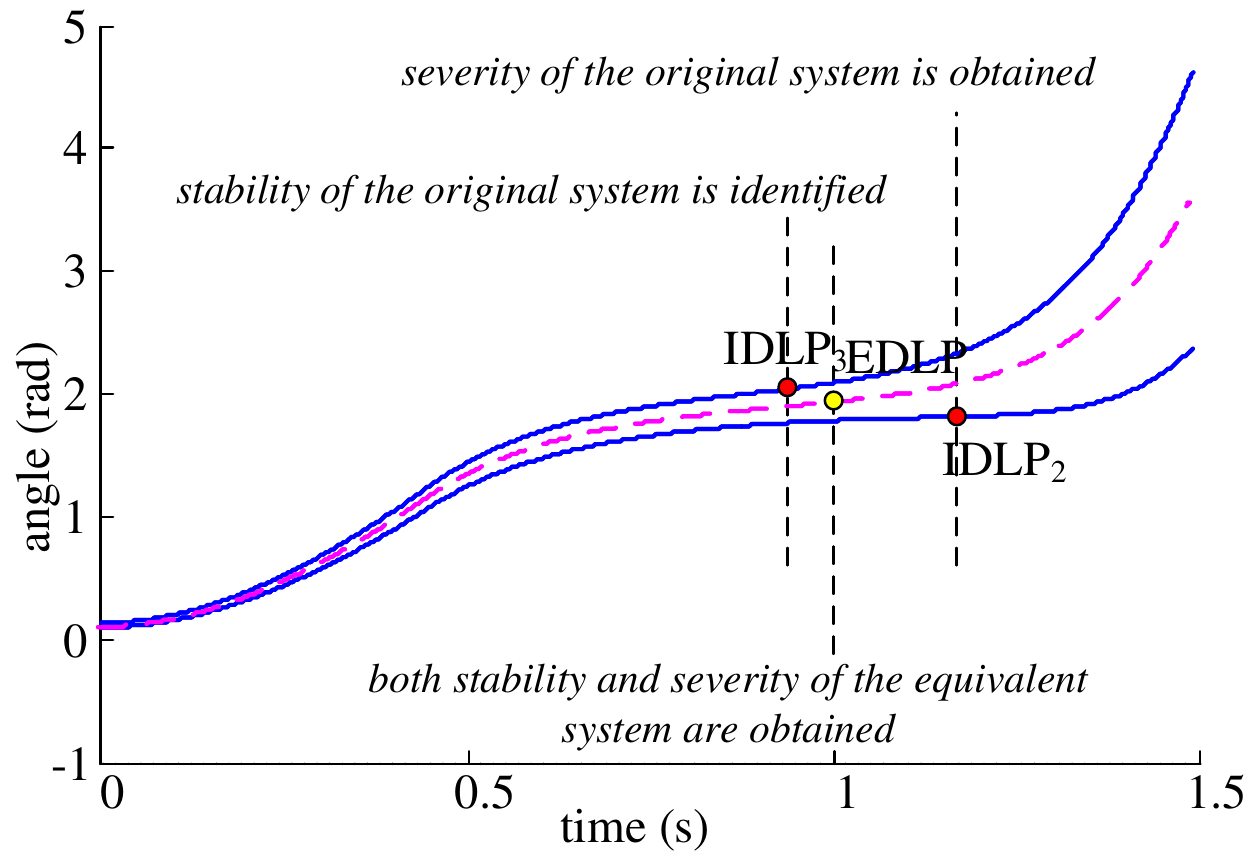}
  \caption{Comparison between the individual machine and the equivalent machine [TS-1, bus-4, 0.447 s].} 
  \label{fig15}  
\end{figure}

 Comparison of the stability evaluation between the original system and the equivalent system is shown in Fig. \ref{fig15}. Detailed analysis was already given in Ref. \cite{3}.
\vspace*{-0.5em}
From Fig. \ref{fig15}, the stability and severity evaluation of the two systems along time horizon are shown as below
\vspace*{0.5em}
\\ \textit{$\text{IDLP}_3$ occurs}: The stability of the original system is characterized as unstable.
\\ \textit{$\text{EDLP}_{\text{CR}}$ occurs}: The stability of the equivalent system is characterized as unstable. The severity of the equivalent system is obtained simultaneously.
\\ \textit{$\text{IDLP}_2$ occurs}: The severity of the original system is obtained.
\vspace*{0.5em}
\par Following the analysis above, the stability evaluations of the equivalent machine show the following characteristics in TSA.
\vspace*{0.5em}
\\ (i) The stability of the equivalent system is characterized ``later'' than that of the original system.
\\ (ii) The severity of the equivalent system is evaluated ``earlier'' than that of the original system.
\vspace*{0.5em}
\par (i) and (ii) fully reflect the ``motion equivalence'' characteristics of the equivalent machine. That is, the equivalent machine shows the “compromise” characteristics between the stability and severity of the system.
\par Essentially speaking, in most simulation cases, all the inner-group machine motions are slight, and thus the original system can be replaced with the equivalent system. Then one can obtain the followings
\\ (i) the severity of the system using the equivalent-machine can be obtained earlier than that using the individual-machine method.
\\ (ii) The precision of the severity evaluation using the equivalent-machine method is also acceptable because all the inner-group machine motions are slight.
\par (i) and (ii) fully indicate that the equivalent-machine is flexible in TSA in most situations when all the inner-group machine motions are slight.

\subsection{``DOUBLE EDGE'' IN THE INDIVIDUAL MACHINE} \label{section_VB}
Compared with the compromise of the equivalent machine, the individual-machine shows different ``double-edged sword'' characteristics in TSA. The two ``double-edged sword'' characteristics are given as below
\vspace*{0.5em}
\\ (i) if the system engineer focuses on the ``efficiency'' of the stability characterization, he has to sacrifice the ``precision'' of the severity evaluation.
\\ (ii) On the contrary, if the system engineer focuses on the ``precision'' of the severity evaluation, he has to tolerate ``efficiency'' of the stability evaluation.
\vspace*{0.5em}
\par (i) and (ii) indicate that the individual-machine has to sacrifice ``one side'' to achieve a better performance of ``the other side''.
\par
\begin{figure}[H]
  \centering
  \includegraphics[width=0.5\textwidth,center]{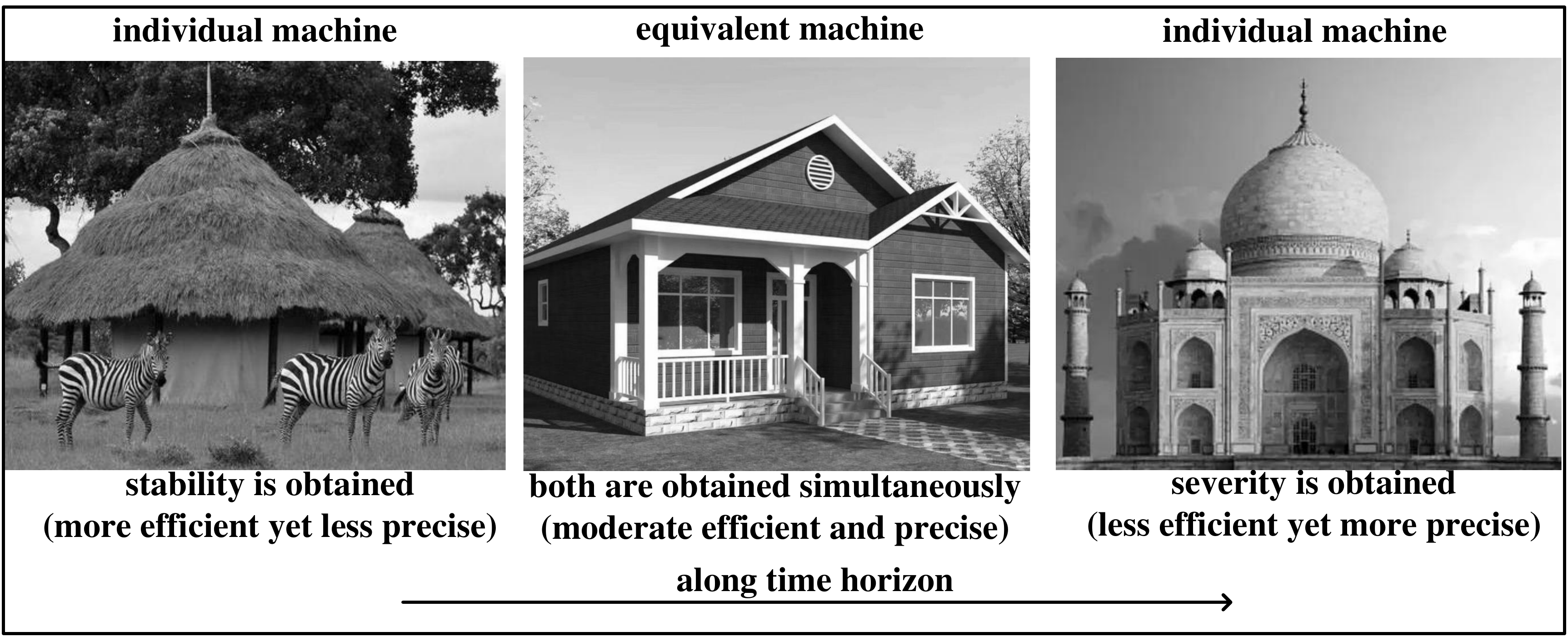}
  \caption{Companion relationship between the equivalent machine method and the individual machine method.} 
  \label{fig16}  
\end{figure}

\begin{figure}[H]
  \centering
  \includegraphics[width=0.45\textwidth,center]{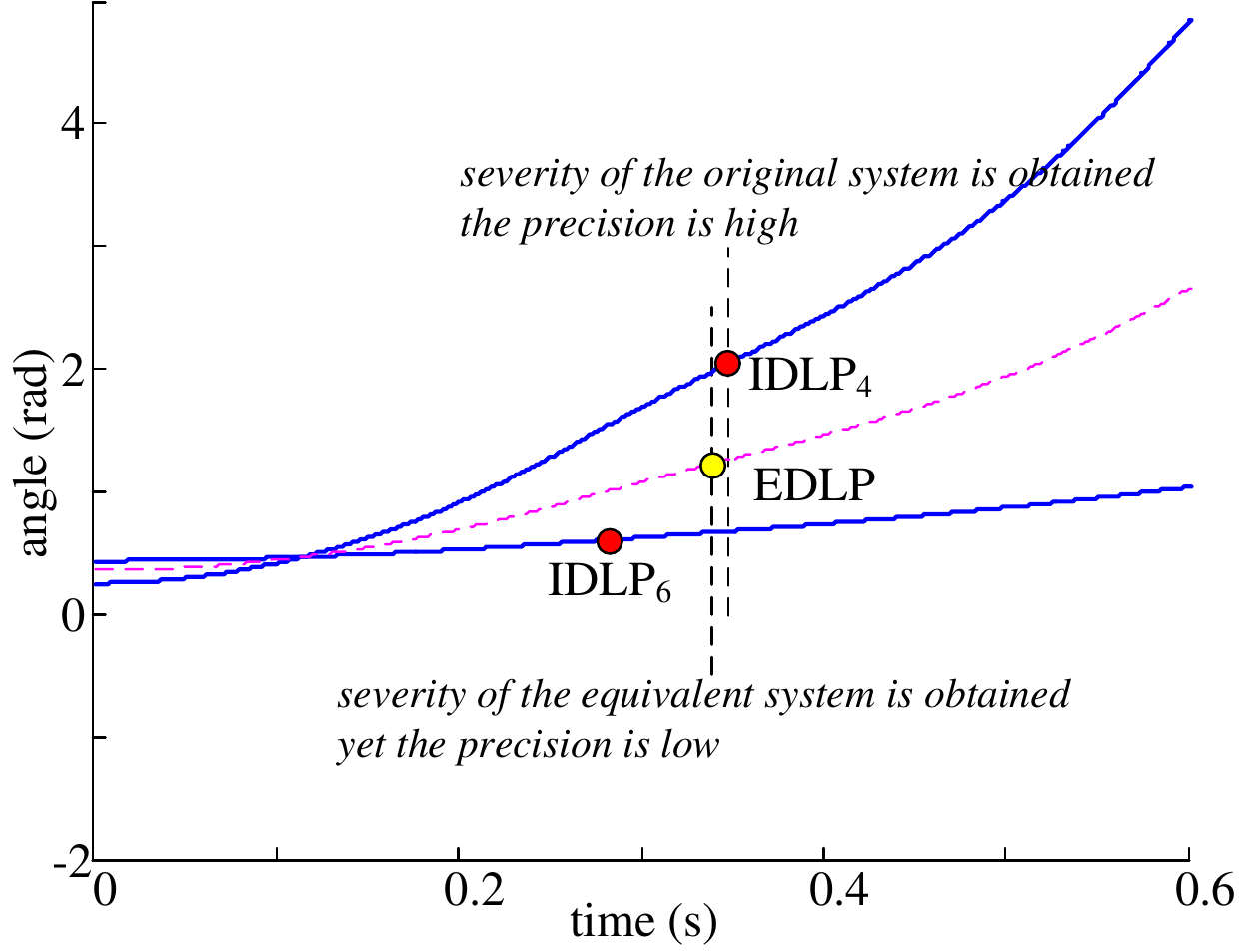}
  \caption{Advantages of the individual-machine in the precise severity evaluation.} 
  \label{fig17}  
\end{figure}

 Based on the analysis above, the ``double edged'' individual-machine can be seen as the ``companion'' of the ``compromised'' equivalent machine method in TSA. This companion relationship is visually shown in Fig. \ref{fig16}.
\vspace*{-0.5em}
Analysis about the “double-edge” characteristics of the individual machine is given as below:
\\ \textit{Particular emphasis in the precision of the severity evaluation}: In actual TSA environment, the system engineer will put a particular emphasis on the ``severity” of the system if the transient stability control is fully considered in TSA.
In order to ensure the precision of the severity evaluation, the severity of the original system can only be evaluated through individual-machine without any equivalence. Under this circumstance, the individual-machine method will show its advantages in the precise severity evaluation in TSA. A tutorial demonstration about this scenario is given in Fig. \ref{fig17}.
\vspace*{-0.5em}
From Fig. \ref{fig17}, using individual machine in TSA, the inner-group machine motion is completely eliminated. The severity of the original system is computed precisely at $\text{IDLP}_4$ by using individual machine. However, note that this severity evaluation is obtained later than that using equivalent machine ($\text{IDLP}_4$ occurs later than $\text{EDLP}_{\text{CR}}$).
\par In brief, the advantage in the ``precision'' of the severity evaluation can be seen as the sacrifice of the ``efficiency'' of the stability characterization when using the individual machine in TSA.
\\ \textit{Particular emphasis in the efficiency of the stability characterization}: if the system engineer only focuses on the ``stability'' of the system, the individual-machine method will also show its advantage with the faster stability characterization along time horizon, as analyzed in Section \ref{section_VA}.
\par This advantage might be reflected in the following scenarios.
\\ (i) The stability characterization rather than the severity evaluation becomes the primary objective of the transient security control.
\\ (ii) The computation of CCT is required.
\vspace*{0.5em}
\par In brief, the advantage in the ``efficiency'' of the stability characterization can be seen as the sacrifice of the ``precision'' of the severity evaluation when using the individual machine in TSA.

\subsection{COMPANION RELATIONSHIP BETWEEN THE TWO MACHINES} \label{section_VC}
Based on all the analysis above, both the equivalent machine and the individual-machine will show advantages in the corresponding scenarios. These advantages under different scenarios are summarized as below in brief
\vspace*{0.5em}
\\ (i) In the TSA environment that all the inner-group machine motions are slight, the equivalent machine will show its advantage with the ``faster'' severity evaluation with acceptable precision.
\\ (ii) In the TSA environment that the precision of the severity evaluation is emphasized (especially when any inner-group machine motion becomes fierce), the individual-machine will show its advantage in the ``precision'' with the sacrifice of the ``efficiency'' of the stability characterization.
\\  (iii) In the TSA environment that the efficiency of the stability characterization is emphasized (especially when the CCT computation is needed), the individual-machine will show its advantage with the sacrifice of the ``efficiency'' of the severity evaluation.
\vspace*{0.5em}
\par (i) to (iii) indicate the followings:
\vspace*{0.5em}
\par \textit{The equivalent machine is flexible in most simulation cases with slight inner-group machine motions}.
\par \textit{Comparatively, the individual-machine can be seen as a ``companion'' of the equivalent method in certain distinctive cases}.
\vspace*{0.5em}
\par Frankly, if equivalent machine is the ``center'' of a football team, then the individual-machine can be seen as the ``forward'' and ``back''.
The individual machine can be used under the distinctive situation especially when inner-group machine motion becomes fierce or the stability characterization becomes the particular emphasis. This can be seen as the guidance for the system engineer to choose a feasible machine under specific TSA environment.

\section{COMPARISON BETWEEN GROUP SEPARATION PATTERN AND MOD} \label{section_VI}
\subsection{DEFINITIONS OF MOD AND GROUP SEPARATION PATTERN} \label{section_VIA}
\noindent \textit{Definition of MOD}: MOD depicts how many machines become unstable for a given disturbance. 
\par Following this definition, assume the system engineer stands at the fault clearing point. At this moment, it is possible that one or more machines might go unstable in future post-fault period.
For a multi-machine system with \textit{n} machines, the motion of machine \textit{n} is not independent in the COI-SYS reference ($\sum_{i=1}^{n} M_{i} \theta_{i}=0$ ). Therefore, the possible combinations of MOD are given as
\begin{equation}
  \label{equ13}
  \sum_{i=1}^{n-1} C_{n-1}^{i}=2^{n-1}-1
\end{equation} 
\par Following Eq. (\ref{equ13}), for a multi-machine system with \textit{n} machines, the possible MODs may reach $2^{n-1}-1$ at the fault clearing point.
\\ \textit{Definition of group separation pattern}: The group separation depicts how many machine form Group-CR.
\par Following this definition, for a multi-machine system with \textit{n} machines, assume one or more machines form $\Omega_{\mathrm{CR}}$. Then the combinations of $\Omega_{\mathrm{CR}}$ can be denoted as
\begin{equation}
  \label{equ14}
  \sum_{i=1}^{n-1} C_{n}^{i}=2^{n}-2
\end{equation}
\par Because $C_{n}^{i}$ and $C_{n}^{n-i}$ corresponds to the same group separation pattern, the real combinations of $\Omega_{\mathrm{cr}}$ would become half of that in Eq. (\ref{equ14}), and thus the combinations of all the possible group separation patterns can be denoted as
\begin{equation}
  \label{equ15}
  \frac{1}{2}\sum_{i=1}^{n-1} C_{n}^{i}=2^{n-1}-1
\end{equation}
\par Eqs. (\ref{equ13}) and (\ref{equ15}) indicate that the combinations of MOD are mathematically equal to the numbers of the possible group separation patterns.

\subsection{CONJECTURE ABOUT THE EQUIVALENCE BETWEEN MOD AND GROUP SEPARATION PATTERN} \label{section_VIB}
We go a further step. Assume a long-time simulation is experienced. Under this circumstance, the final MOD is confirmed, and the final dominant group separation is also identified. It seems that the final MOD is chosen from $2^{n-1}-1$ possible MODs, while the dominant pattern is also chosen from the $2^{n-1}-1$ possible patterns that exist at the fault clearing point.
Based on these ``similarities'', the equivalent machine analysts believed that the dominant group separation pattern is ``identical'' to the real MOD in the theoretical level.
\par
\begin{figure}[H]
  \centering
  \includegraphics[width=0.45\textwidth,center]{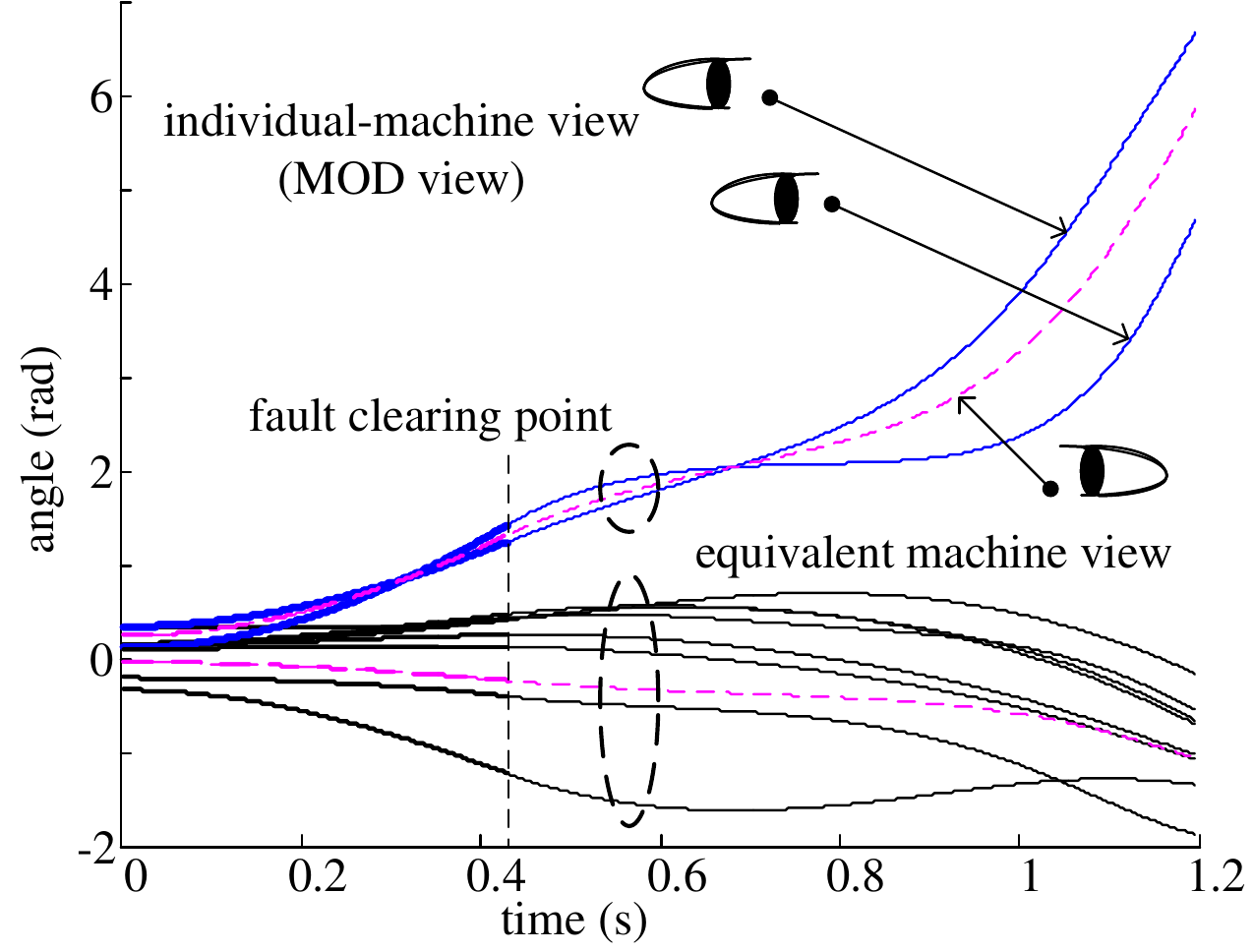}
  \caption{Difference between MOD and dominant group separation pattern [TS-1, bus-2, 0.430 s].} 
  \label{fig18}  
\end{figure}

 A tutorial example is given below. Demonstration about the difference between MOD and dominant group separation pattern is shown in Fig. \ref{fig18}.
\vspace*{-0.5em}
From Fig. \ref{fig18}, assume the system operator is standing at the fault clearing point (0.430s). At the moment he does not know what the system trajectory might become in the following period. The combinations of MOD and that of group separation patterns are $2^9-1$.
After that, at 1.200 s, the real MOD is finally confirmed as Machines 9 and 8 going unstable, while the dominant group separation pattern is defined as Machines 8 and 9 forming $\Omega_{\mathrm{CR}}$.

\subsection{CLARIFICATIONS FROM THE INNER-GROUP MACHINE PERSPECTIVE} \label{section_VIC}
In fact this conjecture is a misunderstanding about the individual machine and the equivalent machine, although the combinations of MOD are mathematically equal to that of the group separation patterns.
\par Based on the definitions of the MOD and group separation pattern, the following deductions can be obtained
\vspace*{0.5em}
\par \textit{The MOD is defined in an individual-machine manner}.
\par \textit{Comparatively, the dominant group separation pattern serves the machine equivalence}.
\vspace*{0.5em}
\par In fact, the difference between MOD and the dominant group separation pattern is just the inner-group machine motion. 
\par Taking the case in Fig. \ref{fig18} as an example, At the first glance, it seems that Machines 9 and 8 are successfully captured inside $\Omega_{\mathrm{CR}}$, and thus MOD is ``identical'' to the group equivalent pattern.
In fact this is a complete misunderstanding. This is because the group separation pattern aims to serve the machine equivalence. In particular, under dominant group separation pattern, the equivalent Machine-CR is finally modeled. Under this circumstance, differences always can be found between the ``non-equivalent'' real machine and the ``equivalent'' Machine-CR, These differences are just the inner-group motions inside $\Omega_{\mathrm{CR}}$.
\par Strictly speaking, the MOD will be close (not identical) to the dominant-group separation pattern if all the inner-group motions are slight, while the MOD will become different from the dominant-group separation pattern if any inner-group machine instability occurs.

\section{REVISITS OF THE MACHINES} \label{section_VII}
In this section, the mechanisms of the individual-machine, superimposed machine, the equivalent machine and the inner-group machine are systematically revisited. All the analysis in this section is based on the COI-SYS reference. This systematic revisit may further help readers take a deep insight into the mechanisms of these machines.
\par The mechanism of the individual machine, superimposed machine and the equivalent machine are shown in Fig. \ref{fig19}.
\begin{figure} [H]
  \centering 
  \subfigure[]{%
  \label{fig19a}
    \includegraphics[width=0.47\textwidth]{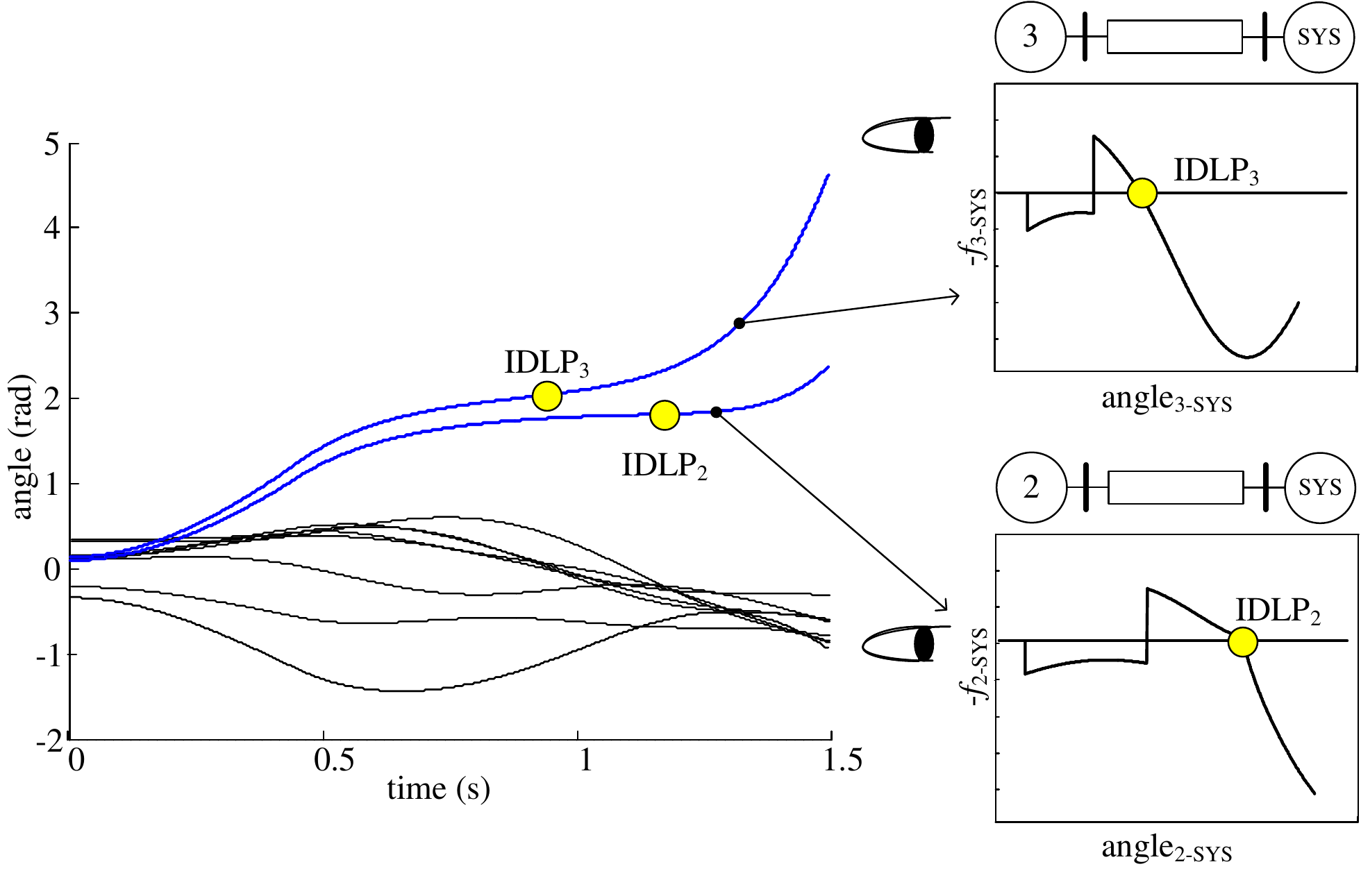}}%
\end{figure} 
\addtocounter{figure}{-1}
\vspace*{-2em}       
\begin{figure} [H]
  \addtocounter{figure}{1}      
  \centering 
  \subfigure[]{%
    \label{fig19b}
    \includegraphics[width=0.47\textwidth]{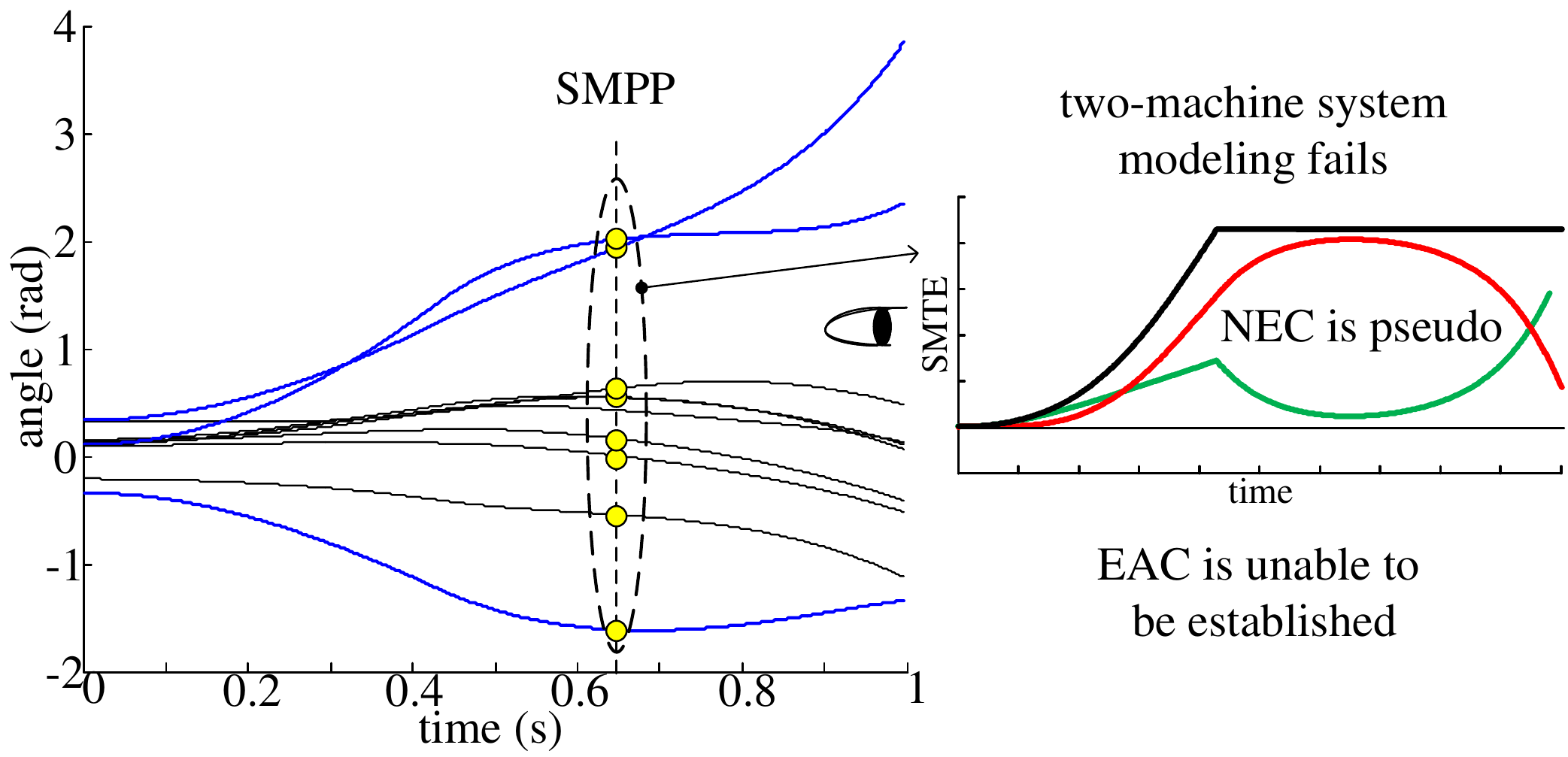}}%
\end{figure}
\addtocounter{figure}{-1}
\vspace*{-2em}       
\begin{figure} [H]
  \addtocounter{figure}{1}      
  \centering 
  \subfigure[]{%
    \label{fig19c}
    \includegraphics[width=0.47\textwidth]{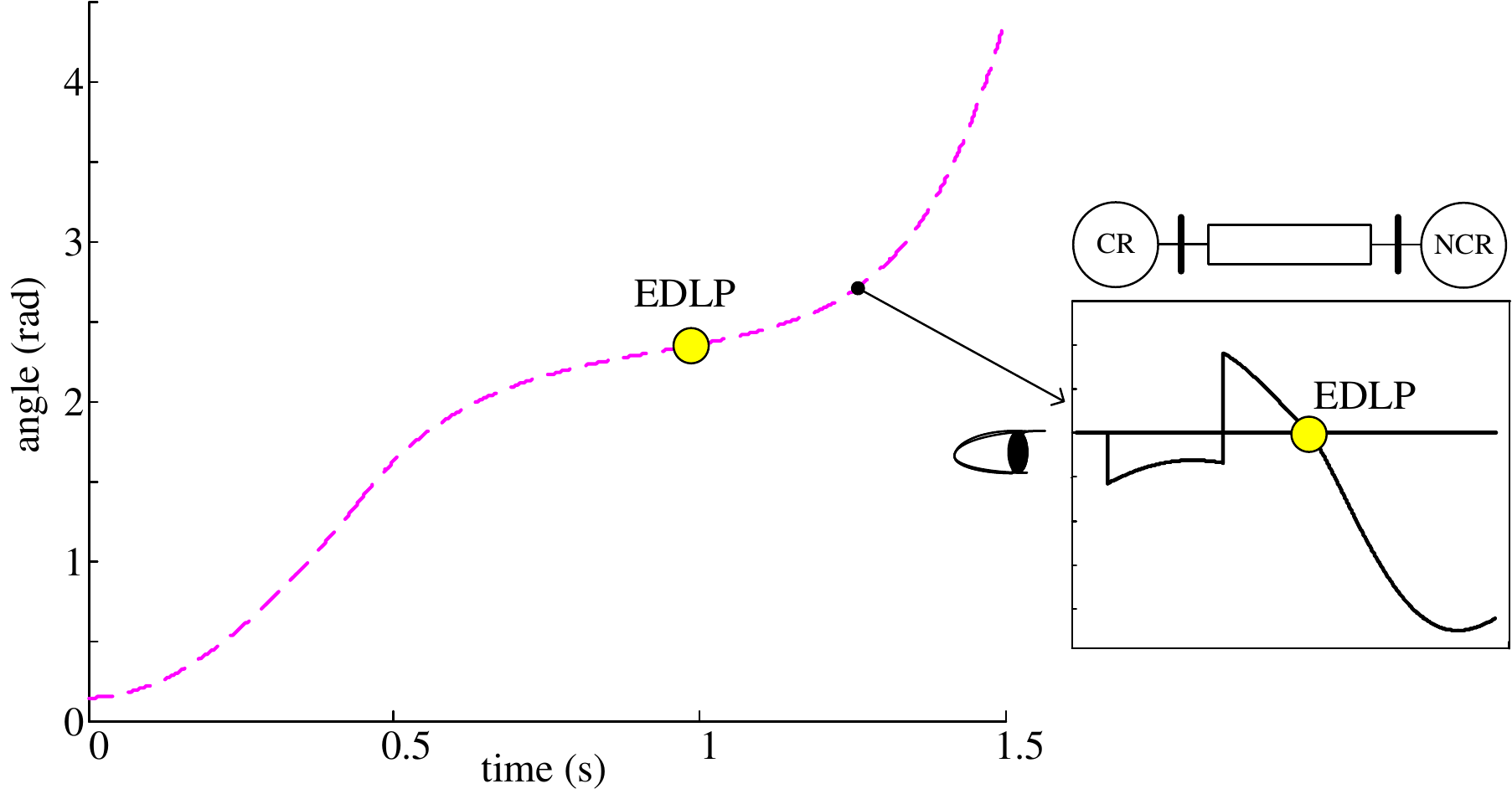}}%
  \caption{Comparison among the four machines. (a) Individual machine. (b) Superimposed machine. (c) Equivalent machine.}%
  \label{fig19}
\end{figure}
\vspace*{-0.5em}
Mechanisms of the three machines are summarized as below
\\
\textit{Individual-machine}: The idea of the machine comes from the ``individual-machine monitoring” of the original system trajectory. The I-SYS system is established through the separation of the individual machine in the COI-SYS reference. Against this background, the IMTE shows strict NEC characteristics. The strict correlation between IMTR and IMTE is established through each I-SYS system.
The stability of the entire original system is evaluated in a ``machine-by-machine'' manner because NEC of each individual machine is unique and different.
\\ \textit{Superimposed machine}: The idea of the machine comes from the ``global monitoring'' of the original system trajectory. However, due to the energy superimposition, the superimposed machine is a pseudo machine without any equation of motion. Against this background, the correlation between the original system trajectory and the SMTE cannot be established, while its transient energy conversion does not satisfy NEC characteristic with always existence of the residual SMKE.
\\ \textit{Equivalent machine}: The idea of the machine comes from the ``equivalent machine monitoring'' through group separation and motion equivalence.
The CR-NCR system is established through the separation of Machine-CR in the COI-NCR reference. Against this background, Machine-CR becomes the ``one-and-only'' machine in TSA. The EMTE shows strict NEC characteristics. The strict correlation between EMTR and EMTE is established through CR-NCR system. The stability of the equivalent system is evaluated in a ``one-and-only machine'' manner.
\par From analysis above, the mistakenly global monitoring leads to the energy superimposition in the superimposed machine. This energy superimposition causes the superimposed machine to become a pseudo machine without any equation of motion. Against this background, the pseudo superimposed machine completely violates all the machine paradigms and it show inherit defects in the TSA. Comparatively, the equivalent machine is modeled based on the motion equivalence of all individual machines inside each group. This motion equivalence ensures the equivalent machine to have its equivalent equation of motion. Against this background, the equivalent machine strictly follows machine paradigms and it show advantages in TSA. Note that the inner-group machine is created from the ``difference'' between the individual machine and equivalent machine.
\par In actual TSA environments, the individual machine and equivalent machine can be seen as ``companions''. To be specific, the equivalent machine will show flexibility when the equivalent system is close to the original system, while the individual-machine will become useful in the distinctive environment that especially emphasizes the ``more efficient stability characterization'' or ``more precise severity evaluation''.

\section{CONCLUSIONS} \label{section_VIII}
This paper focuses on the detailed modeling and stability characterizations of the inner-group motions. The inner-group motion it is created from the difference between the equivalent system and the original system. It physically does not exist. Following transient stability paradigms, the trajectory variance of the inner-group machine can be modeled through the I-CR system with strict NEC characteristic. The transient characteristics of the inner-group machine are analyzed. The inner-group motions might be inseparable or separable. The inner-group instability will occur when the inner-group machine separates from Machine-CR. The companion relationships between the individual-machine and the equivalent machine are analyzed. It is found that the equivalent machine method show flexibility when the inner-group motions are slight.
Comparatively, the individual-machine will become useful under certain distinctive situations in which ``more efficient stability characterization'' or ``more precise severity evaluation'' becomes a particular emphasis. In the end of the paper, it is clarified that MOD will become significantly different from the dominant-group separation pattern if any inner-group machine instability occurs.
A systematical revisit about the individual-machine, superimposed machine and the equivalent machine is provided. This revisit essentially enhances the understandings of the mechanisms of the proposed machines in the previous papers.
\par In the history of the power system transient stability, quite different from the original definition of the equivalent machine that is modeled based on the motion equivalence, the equivalent machine analysts also attempt to analyze the ``machine transformation'' from the individual machine to the equivalent machine from the perspective the ``correction'' of the inner-group machines. This will be analyzed in the following paper.

%

%
%
%





\begin{thebibliography}{1}
\bibitem{1}
S. Wang, J. Yu, and A. Foley, ``Newtonian Mechanics Based Transient Stability PART I: Transient Stability Paradigms''.

\bibitem{2}
S. Wang, J. Yu, and A. Foley, ``Newtonian Mechanics Based Transient Stability PART II: Individual Machine''.

\bibitem{3}
S. Wang, J. Yu, and A. Foley, ``Newtonian Mechanics Based Transient Stability PART IV: Equivalent Machine''.

\bibitem{4}
D. Fang, T. S. Chung, Y. Zhang, and W. Song, ``Transient stability limit conditions analysis using a corrected transient energy function approach``', IEEE Trans. Power Syst. vol. 15, no. 2, pp. 804-810, 2000.

\bibitem{5}
Y. Xue, ``Re-examination of transient energy functions and critical energy'', Automation of Electric Power Systems, vol. 6, pp. 9-18, 1991.

\end{thebibliography}
\end{document}